\documentclass[12pt]{article}

\makeatletter

\def\section{\@startsection {section}{1}{\z@}{+3.0ex plus +1ex minus
  +.2ex}{2.3ex plus .2ex}{\normalsize\bf}}
\def\subsection{\@startsection{subsection}{2}{\z@}{+2.5ex plus +1ex
minus +.2ex}{1.5ex plus .2ex}{\normalsize\bf}}
\def\subsubsection{\@startsection{subsubsection}{3}{\z@}{+3.25ex plus
 +1ex minus +.2ex}{1.5ex plus .2ex}{\normalsize\bf}}

\expandafter\ifx\csname mathrm\endcsname\relax\def\mathrm#1{{\rm #1}}\fi

\makeatletter

\newcount\@tempcntc
\def\@citex[#1]#2{\if@filesw\immediate\write\@auxout{\string\citation{#2}}\fi
  \@tempcnta\z@\@tempcntb\m@ne\def\@citea{}\@cite{\@for\@citeb:=#2\do
    {\@ifundefined
       {b@\@citeb}{\@citeo\@tempcntb\m@ne\@citea
        \def\@citea{,\penalty\@m\ }{\bf ?}\@warning
       {Citation `\@citeb' on page \thepage \space undefined}}%
    {\setbox\z@\hbox{\global\@tempcntc0\csname
b@\@citeb\endcsname\relax}%
     \ifnum\@tempcntc=\z@ \@citeo\@tempcntb\m@ne
       \@citea\def\@citea{,\penalty\@m}
       \hbox{\csname b@\@citeb\endcsname}%
     \else
      \advance\@tempcntb\@ne
      \ifnum\@tempcntb=\@tempcntc
      \else\advance\@tempcntb\m@ne\@citeo
      \@tempcnta\@tempcntc\@tempcntb\@tempcntc\fi\fi}}\@citeo}{#1}}

\def\@citeo{\ifnum\@tempcnta>\@tempcntb\else\@citea
  \def\@citea{,\penalty\@m}%
  \ifnum\@tempcnta=\@tempcntb\the\@tempcnta\else
   {\advance\@tempcnta\@ne\ifnum\@tempcnta=\@tempcntb \else
\def\@citea{--}\fi
    \advance\@tempcnta\m@ne\the\@tempcnta\@citea\the\@tempcntb}\fi\fi}

\def\nl{\nonumber\\}

\newcommand{\lsim}
{\mathrel{\raisebox{-.3em}{$\stackrel{\displaystyle <}{\sim}$}}}

\def\asymp#1%
{\mathrel{\raisebox{-.4em}{$\widetilde{\scriptstyle #1}$}}}

\def\Nequal#1%
{\mathrel{\raisebox{-.5em}{$\stackrel{=}{\scriptstyle\rm#1}$}}}

\def\beq{\begin{equation}}
\def\eeq{\end{equation}}
\def\beqar{\begin{eqnarray}}
\def\eeqar{\end{eqnarray}}
\def\barr#1{\begin{array}{#1}}
\def\earr{\end{array}}
\def\bfi{\begin{figure}}
\def\efi{\end{figure}}
\def\btab{\begin{table}}
\def\etab{\end{table}}
\def\bce{\begin{center}}
\def\ece{\end{center}}
\def\nn{\nonumber}

\def\text{\textstyle}

\def\al{\alpha}
\def\be{\beta}
\def\Ga{\Gamma}
\def\ga{\gamma}
\def\de{\delta}
\def\De{\Delta}
\def\eps{\epsilon}
\def\veps{\varepsilon}
\def\la{\lambda}

\def\Om{\Omega}
\def\si{\sigma}

\def\refeq#1{\mbox{(\ref{#1})}}

\def\reffi#1{\mbox{Fig.~\ref{#1}}}
\def\reffis#1{\mbox{Figs.~\ref{#1}}}
\def\refta#1{\mbox{Table~\ref{#1}}}

\def\refse#1{\mbox{Section~\ref{#1}}}
\def\refapp#1{\mbox{App.~\ref{#1}}}
\def\citere#1{\mbox{Ref.~\cite{#1}}}

\def\solid{\raise.9mm\hbox{\protect\rule{1.1cm}{.2mm}}}
\def\dash{\raise.9mm\hbox{\protect\rule{2mm}{.2mm}}\hspace*{1mm}}
\def\dot{\rlap{$\cdot$}\hspace*{2mm}}
\def\dashed{\dash\dash\dash\dash}

\def\dashdotted{\dot\dash\dot\dash\dot}

\newcommand{\TeV}{\unskip\,\mathrm{TeV}}
\newcommand{\GeV}{\unskip\,\mathrm{GeV}}
\newcommand{\MeV}{\unskip\,\mathrm{MeV}}
\newcommand{\pb}{\unskip\,\mathrm{pb}}
\newcommand{\fb}{\unskip\,\mathrm{fb}}

\newcommand{\rd}{{\mathrm{d}}}

\newcommand{\Oa}{\mathswitch{{\cal{O}}(\alpha)}}

\newcommand{\M}{{\cal{M}}}

\def\mathswitchr#1{\relax\ifmmode{\mathrm{#1}}\else$\mathrm{#1}$\fi}
\newcommand{\Pf}{\mathswitch  f}

\newcommand{\PW}{\mathswitchr W}
\newcommand{\PZ}{\mathswitchr Z}

\newcommand{\PH}{\mathswitchr H}
\newcommand{\Pe}{\mathswitchr e}

\newcommand{\Pd}{\mathswitchr d}
\newcommand{\Pu}{\mathswitchr u}
\newcommand{\Ps}{\mathswitchr s}
\newcommand{\Pc}{\mathswitchr c}
\newcommand{\Pb}{\mathswitchr b}
\newcommand{\Pbbar}{\mathswitchr{\bar b}}
\newcommand{\Pt}{\mathswitchr t}
\newcommand{\Ptbar}{\mathswitchr{\bar t}}
\newcommand{\Pep}{\mathswitchr {e^+}}
\newcommand{\Pem}{\mathswitchr {e^-}}
\newcommand{\PWp}{\mathswitchr {W^+}}
\newcommand{\PWm}{\mathswitchr {W^-}}

\def\mathswitch#1{\relax\ifmmode#1\else$#1$\fi}
\newcommand{\Mf}{\mathswitch {m_\Pf}}

\newcommand{\MW}{\mathswitch {M_\PW}}

\newcommand{\MZ}{\mathswitch {M_\PZ}}
\newcommand{\MH}{\mathswitch {M_\PH}}
\newcommand{\Me}{\mathswitch {m_\Pe}}

\newcommand{\Md}{\mathswitch {m_\Pd}}
\newcommand{\Mu}{\mathswitch {m_\Pu}}
\newcommand{\Ms}{\mathswitch {m_\Ps}}
\newcommand{\Mc}{\mathswitch {m_\Pc}}
\newcommand{\Mb}{\mathswitch {m_\Pb}}
\newcommand{\Mt}{\mathswitch {m_\Pt}}

\newcommand{\sw}{\mathswitch {s_\PW}}
\newcommand{\cw}{\mathswitch {c_\PW}}

\newcommand{\GF}{\mathswitch {G_\mu}}
\newcommand{\VA}{\mathrm{\{V,A\}}}
\newcommand{\V}{\mathrm{V}}

\def\Li{\mathop{\mathrm{Li}}\nolimits}
\def\Re{\mathop{\mathrm{Re}}\nolimits}
\def\ie{i.e.\ }

\newcommand{\se}{self-energy}

\newcommand{\cs}{cross-section}
\newcommand{\css}{cross-sections}

\newcommand{\AAWW}{\gamma\gamma\to\PWp\PWm}
\newcommand{\AAtt}{\gamma\gamma\to\Pt\Ptbar}

\newcommand{\Born}{\mathrm{Born}}

\newcommand{\born}{\mathrm{Born}}

\newcommand{\weak}{\mathrm{weak}}
\newcommand{\bos}{\mathrm{bos}}
\newcommand{\ferm}{\mathrm{ferm}}
\newcommand{\Coul}{\mathrm{Coul.}}

\newcommand{\cut}{\mathrm{cut}}
\newcommand{\SB}{\mathrm{SB}}

\newcommand{\CMS}{\mathrm{CMS}}
\renewcommand{\CMS}{\mathrm{}}
\newcommand{\QED}{\mathrm{QED}}

\renewcommand{\min}{\mathrm{min}}
\renewcommand{\max}{\mathrm{max}}

\newcommand{\dsl}[1]{\not \hspace{-0.7mm}#1}
\def\dsl{\mathpalette\make@slash}
\def\make@slash#1#2{\setbox\z@\hbox{$#1#2$}%
  \hbox to 0pt{\hss$#1/$\hss\kern-\wd0}\box0}

\newcommand{\AAZ}{$AAZ$}
\newcommand{\AAX}{$AA\chi$}

\def\draftdate{\relax}
\def\mda{\relax}
\def\mua{\relax}
\def\mla{\relax}
\def\draft{
\def\thtystars{******************************}
\def\sixtystars{\thtystars\thtystars}
\typeout{}
\typeout{\sixtystars**}
\typeout{* Draft mode!
         For final version remove \protect\draft\space in source file *}
\typeout{\sixtystars**}
\typeout{}
\def\draftdate{\today}
\def\mua{\marginpar[\boldmath\hfil$\uparrow$]%
                   {\boldmath$\uparrow$\hfil}%
                    \typeout{marginpar: $\uparrow$}\ignorespaces}
\def\mda{\marginpar[\boldmath\hfil$\downarrow$]%
                   {\boldmath$\downarrow$\hfil}%
                    \typeout{marginpar: $\downarrow$}\ignorespaces}
\def\mla{\marginpar[\boldmath\hfil$\rightarrow$]%
                   {\boldmath$\leftarrow $\hfil}%
                    \typeout{marginpar: $\leftrightarrow$}\ignorespaces}
\overfullrule 5pt
\oddsidemargin -15mm
\marginparwidth 29mm
}

\def\stars{\strut\leaders\hbox{*}\hfill\strut}
\def\starline{\hfil\strut\hfil\hbox to \textwidth {\stars}\hfil}

\begin{document}

\thispagestyle{empty}
\def\thefootnote{\fnsymbol{footnote}}
\setcounter{footnote}{1}
\null
\strut\hfill  BI-TP 95/27 \\
\strut\hfill WUE-ITP-95-017\\
\strut\hfill hep-ph/9507372
\vskip 0cm
\vfill
\begin{center}
{\Large \bf
\boldmath{Radiative Corrections to $\gamma\gamma\to\Pt\Ptbar$ \\
    in the Electroweak Standard Model}
\par} \vskip 2.5em
{\large
{\sc A.~Denner%
}\\[1ex]
{\normalsize \it Institut f\"ur Theoretische Physik, Universit\"at W\"urzburg\\
D-97074 W\"urzburg, Am Hubland, Germany}
\\[2ex]
{\sc S.~Dittmaier%
\footnote{Partially supported by the Bundesministerium f\"ur Bildung und
Forschung, Bonn, Germany.} }\\[1ex]
{\normalsize \it Theoretische Physik, Universit\"at Bielefeld\\
D-33501 Bielefeld, Postfach 100131, Germany}
\\[2ex]
{\sc M.~Strobel%
} \\[1ex]
{\normalsize \it Institut f\"ur Theoretische Physik, Universit\"at W\"urzburg\\
D-97074 W\"urzburg, Am Hubland, Germany}%
\vfill
}%
\par \vskip 1em%
\end{center}\par
\vskip 1cm
{\bf Abstract:} \par
The \cs\ for $\gamma\gamma\to\Pt\Ptbar$ with arbitrary
polarized photons is calculated within the electroweak
Standard Model including the complete virtual and soft-photonic
${\cal O}(\alpha)$ corrections. We present a detailed numerical discussion
of the radiative corrections with particular emphasis on the
purely weak corrections.
These are usually of the order of
1--10\% for energies up to 1 TeV.
For unpolarized or equally polarized photons they
reach almost 10\% close to threshold.
The large corrections cannot be traced back to a
universal origin like the running of $\alpha$ or the $\rho$-parameter.
Apart from the energy region around the Higgs resonance
$(\gamma\gamma\to\PH^*\to\Pt\Ptbar)$ the weak corrections are widely
independent of the Higgs-boson mass.
\par
\vskip 1cm
\noindent BI-TP 95/27 \\
WUE-ITP-95-017\par
\vskip .15mm
\noindent July 1995 \par
\null
\setcounter{page}{0}
\clearpage
\def\thefootnote{\arabic{footnote}}
\setcounter{footnote}{0}

\section{Introduction}
\label{se:intro}

The recent discovery of the top quark at the Tevatron \cite{mtcdf,mtd0}
was a further important step in establishing the Standard Model of the
electroweak interaction (SM).
The experimental values for the top-quark mass
$\Mt=176 \pm 8 \pm 10\GeV$ from CDF\cite{mtcdf} and
$\Mt=199^{+19}_{-21} \pm 22\GeV$ from D\O\cite{mtd0}
have been found to be in very good agreement with the SM prediction of
$\Mt=173^{+12\,+18}_{-13\,-20} \GeV$
based on precision tests at LEP1 \cite{mtlep}.
A precision measurement of the top-quark mass together with the accurate
value for the \PW-boson mass to be expected from LEP2 will allow to
derive indirect limits on the mass of the SM Higgs boson.
Moreover, many properties of the top quark such as
its width and its couplings still remain to be studied experimentally.

For precision measurements of the properties of the top quark,
electron-positron colliders are much better suited than proton--(anti-)%
proton colliders \cite{Be92}. The analysis of the process
$\Pep\Pem\to\Pt\bar\Pt$ will allow to determine the top-quark
mass at the level of 0.3\%, the magnetic dipole moments of the top quark
at the few-percent level, and even the Higgs Yukawa coupling to the top
quark.

Using Compton backscattering of laser photons off high-energy electrons,
a $\Pep\Pem$ collider can be turned into a $\ga\ga$ collider
\cite{nlc}.
This mode provides a rich variety of interesting physical processes
(for a review see e.g.\ \citere{Br95}). In particular, it
allows complementary investigations of the top quark by
using the reaction $\AAtt$. Similar to the process
$\Pep\Pem\to\Pt\bar\Pt$, an accurate investigation of its threshold
allows to extract $\Mt$ and $\alpha_s$ and even $\Gamma_\Pt$
\cite{Bi93}. At high energies, the corresponding cross-section
is larger than the one of $\Pep\Pem\to\Pt\bar\Pt$.
The reaction $\AAtt$ is
particularly suited to study the top-quark--photon coupling \cite{Ch95}.

At a center-of-mass energy of $500\GeV$ the cross-section
for $\AAtt$ amounts to $\si\sim 0.8\pb$ and thus,
assuming an integrated luminosity of $10 \fb^{-1}$,
gives rise to about 8000 events,
corresponding to a statistical error of roughly 1\%. At this level of
experimental accuracy an
investigation of the one-loop radiative corrections is in order.
The one-loop QCD corrections have already been studied in \citere{Ku93}
and found to be large.
Since the top quarks can only be detected via their decay products,
e.g.\ in $\Pt\to\Pb\PWp$, finite-widths effects of the top quark and
irreducible background contributions, e.g.\
$\ga\ga\to\Pb\Pbbar\PWp\PWm$, have to be taken into account. At tree
level these contributions have been investigated in \citere{mo95}.

As far as the electroweak one-loop corrections are concerned, so far
only the Higgs-dependent corrections have been calculated \cite{Bo92}
and the Higgs resonance contributions have been investigated
\cite{Ve94}.

In this paper we present the results of a calculation of the complete
virtual and soft-photonic \Oa\ radiative corrections to $\AAtt$ in the
SM.
The calculation was carried out with help of {\it Mathematica}
\cite{math}.
The Feynman graphs were generated and drawn by {\it FeynArts} \cite{fa}.
We performed two independent calculations, one using {\it FeynCalc}
\cite{fc} and one using an independent package for loop calculations.
As the analytical result is very lenghty and untransparent, we refrain
from writing it down in full detail. We indicate its general
structure and give analytical results only for the most important $\Oa$
corrections.
We discuss in particular the fermion-loop corrections,
the leading effects from light fermions,
the Coulomb singularity, the Higgs resonance, and the heavy-Higgs effects.
In the discussion of the numerical results of course the complete
corrections are included. While we have calculated these corrections
for arbitrary polarized photons and top quarks, we restrict the
presentation to unpolarized top quarks in this paper.

The paper is organized as follows: The notation and conventions are
given in \refse{se:notcon}. Section \ref{se:born} summarizes the
lowest-order results. The evaluation and general features of the
higher-order corrections are discussed in \refse{se:RC}, the
numerical results in \refse{se:numres}. In the appendix we list
some
explicit formulae.

\section{Notation and conventions}
\label{se:notcon}

We consider the reaction
\begin{equation}
\gamma(k_{1},\lambda_{1}) + \gamma(k_{2},\lambda_{2}) \; \longrightarrow \;
\mbox{t}(p,\sigma) + \bar{\mbox{t}}(\bar{p},\bar{\sigma}),
\end{equation}
where $\lambda_{1,2} = \pm 1$ denote the helicities of the incoming photons
and $\sigma,\bar{\sigma} = \pm 1/2$ the spin
orientations of the outgoing top quarks.

In the center-of-mass system (CMS) the momenta read
in terms of the energy $E$ of the incoming photons and
the scattering angle $\theta$
\begin{eqnarray}
k_{1}^{\mu}   & = & E(1,0,0,-1), \nonumber \\
k_{2}^{\mu}   & = & E(1,0,0,1),  \nonumber \\
p^{\mu}       & = & E(1,-\beta \sin{\theta},0,-\beta \cos{\theta}), \nonumber
\\
\bar{p}^{\mu} & = & E(1,\beta \sin{\theta},0,\beta \cos{\theta}),
\label{momenta}
\end{eqnarray}
where $\beta = \sqrt{1-\Mt^2/E^2}$ is the velocity of the top quarks in
the CMS.
The Mandelstam variables are given by
\begin{eqnarray}
s & = & (k_{1} + k_{2})^{2}   \; =  \; (p + \bar{p})^{2}
 \phantom{_{2}} \; = \; 4 E^{2},   \nonumber \\
t & = & (k_{1}-p)^{2} \phantom{_{2}}  \;  =  \; (k_{2} - \bar{p})^{2}
  \; = \; \Mt^{2} - \frac{s}{2}(1 - \beta \cos{\theta}), \nonumber \\
u & = & (k_{1} - \bar{p})^{2} \phantom{_{2}} \; =  \; (k_{2} - p)^{2}
  \; = \; \Mt^{2} - \frac{s}{2}(1 + \beta \cos{\theta}).
\label{stu}
\end{eqnarray}

In order to calculate polarized \css, we introduce explicit polarization
vectors for the photons as follows
\begin{eqnarray}
\veps_{1}^{\mu}(k_{1},\lambda_{1} = \pm 1) & = &
  -\frac{1}{\sqrt{2}}(0,1,\mp \:i,0), \nonumber \\
\veps_{2}^{\mu}(k_{2},\lambda_{2} = \pm 1) & = & \phantom{-}\frac{1}{\sqrt{2}}
  (0,1,\pm \: i,0).
\label{eq:poldef}
\end{eqnarray}

The scattering amplitude of $\gamma\gamma\to\Pt\Ptbar$ obeys Bose symmetry
with respect to the incoming photons
and---neglecting quark mixing---%
also CP symmetry. Consequently, the helicity amplitudes
$\M_{\la_1\la_2\si\bar{\si}}$ are related by
\begin{eqnarray}
\M_{\la_1\la_2\si\bar{\si}}(s,t,u) &=&
\parbox{4.5cm}{$\M_{\la_2\la_1\si\bar{\si}}(s,u,t)$}
(\mbox{Bose}), \nn\\
\M_{\la_1\la_2\si\bar{\si}}(s,t,u) &=&
\parbox{4.5cm}{$\M_{-\la_1-\la_2-\bar\si-\si}(s,u,t)$}
(\mbox{CP}), \nn\\
\M_{\la_1\la_2\si\bar{\si}}(s,t,u) &=&
\parbox{4.5cm}{$\M_{-\la_2-\la_1-\bar\si-\si}(s,t,u)$}
(\mbox{Bose + CP}).
\label{eq:BoseCP}
\end{eqnarray}

In the following we restrict ourselves to unpolarized top quarks.
In this case, Bose and CP symmetry imply for the differential
\css\ $(\rd\si^{\la_1\la_2}/\rd\Om_\CMS)$
\beqar
\left(\frac{\rd\si^{--}}{\rd\Om_\CMS}\right)(s,t,u) &=&
\left(\frac{\rd\si^{++}}{\rd\Om_\CMS}\right)(s,t,u) \;=\;
\left(\frac{\rd\si^{\pm\pm}}{\rd\Om_\CMS}\right)(s,u,t), \nn\\[.5em]
\left(\frac{\rd\si^{+-}}{\rd\Om_\CMS}\right)(s,t,u) &=&
\left(\frac{\rd\si^{-+}}{\rd\Om_\CMS}\right)(s,u,t).
\label{eq:BoseCP2}
\eeqar
This means that the \css\ for equally polarized photons $(\pm\pm)$ are
equal and forward--backward-symmetric, and the ones for oppositely polarized
photons $(\pm\mp)$ transform into each other via $\cos\theta\to-\cos\theta$.

In lowest order $\gamma\gamma\to\Pt\Ptbar$ is a pure QED process
and therefore invariant under parity P.
Hence, Born
amplitudes and Born \css\ obey the additional relations
\beq
\M_{\Born,\la_1\la_2\si\bar{\si}}(s,t,u) \;=\;
\M_{\Born,-\la_1-\la_2-\si-\bar{\si}}(s,t,u) \qquad\quad
(\mbox{P}),
\eeq
and
\beq
\left(\frac{\rd\si^{+-}_\Born}{\rd\Om_\CMS}\right)(s,t,u) \;=\;
\left(\frac{\rd\si^{-+}_\Born}{\rd\Om_\CMS}\right)(s,t,u),
\eeq
respectively.

The scattering amplitude ${\cal M}$ can be decomposed into invariant functions
$F_{i}^{\VA,t}$ and standard matrix elements (SME)
$\M_{i}^{\VA,t}$, which contain the whole
information about the photon polarizations
\begin{equation}
\M = \sum_i F^{\rm V}_i\M^{\rm V}_i + \sum_i F^{\rm A}_i\M^{\rm A}_i.
\label{eq:sum1}
\end{equation}
Assuming CP invariance and following \citere{ggtt}, the SME can be chosen as
\begin{eqnarray}
\M_1^{\VA,t} &=& \bar{u}(p)
	\, \dsl{\veps}_1(\dsl{\bar{p}}-\dsl{k}_2)\dsl{\veps}_2
	\, \{1,\gamma_5\} \, v(\bar{p}), \nn\\
\M_2^{\VA,t} &=& \bar{u}(p)
	\, (\dsl{k}_1-\dsl{k}_2)
   	\, \{1,\gamma_5\} \, v(\bar{p})
	\, (\veps_1\cdot\veps_2), \nn\\
\M_3^{\VA,t} &=& \bar{u}(p)
	\, [\dsl{\veps}_1 (\veps_2 k_1)-\dsl{\veps}_2 (\veps_1 k_2)]
	\, \{1,\gamma_5\} \, v(\bar{p}), \nn\\
\M_4^{\VA,t} &=& \bar{u}(p)
	\, [\dsl{\veps}_1 (\veps_2\bar{p})-\dsl{\veps}_2(\veps_1 p)]
	\, \{1,\gamma_5 \} \, v(\bar{p}), \nn\\
\M_5^{\VA,t} &=& \bar{u}(p)
	\, (\dsl{k}_1-\dsl{k}_2)
	\, \{1,\gamma_5 \} \, v(\bar{p})
	\, (\veps_1\cdot k_2)(\veps_2\cdot k_1), \nn\\
\M_6^{\VA,t} &=& \bar{u}(p)
	\, (\dsl{k}_1-\dsl{k}_2)
	\, \{1,\gamma_5 \} \, v(\bar{p})
	\, (\veps_1\cdot p)(\veps_2\cdot\bar{p}), \nn\\
\M_7^{\VA,t} &=& \bar{u}(p)
	\, ( \dsl{k}_1-\dsl{k}_2)
	\, \{1,\gamma_5\} \, v(\bar{p})
	\, [(\veps_1\cdot k_2)(\veps_2\cdot\bar{p})
          +(\veps_1\cdot p)(\veps_2\cdot k_1)], \nn\\[2ex]
\M_{11}^{\V,t} &=& \Mt \, \bar{u}(p)
	\, \dsl{\veps}_1\dsl{\veps}_2
	\, v(\bar{p}), \nn\\
\M_{12}^{\V,t} &=& \Mt \, \bar{u}(p)
	\, v(\bar{p})
	\, (\veps_1\cdot\veps_2), \nn\\
\M_{13}^{\VA,t} &=& \Mt \, \bar{u}(p)
	\, [\dsl{\veps}_1\dsl{k}_1(\veps_2 k_1) \pm
	    \dsl{k}_2\dsl{\veps}_2(\veps_1 k_2)]
	\, \{1,\gamma_5\} \, v(\bar{p}), \nn\\
\M_{14}^{\VA,t} &=& \Mt \, \bar{u}(p)
	\, [\dsl{\veps}_1\dsl{k}_1(\veps_2\bar{p}) \pm
	    \dsl{k}_2\dsl{\veps}_2(\veps_1 p)]
	\, \{1,\gamma_5\} \, v(\bar{p}), \nn\\
\M_{15}^{\V,t} &=& \Mt \, \bar{u}(p)
   	\, v(\bar{p})
	\, (\veps_1\cdot k_2)(\veps_2\cdot k_1), \nn\\
\M_{16}^{\V,t} &=& \Mt \, \bar{u}(p)
	\, v(\bar{p})
	\, (\veps_1\cdot p)(\veps_2\cdot \bar{p}), \nn\\
\M_{17}^{\VA,t} &=& \Mt \, \bar{u}(p)
	\, \{1,\gamma_5\} \, v(\bar{p})
	\, [(\veps_1\cdot k_2)(\veps_2\cdot \bar{p}) \pm
            (\veps_1\cdot p)(\veps_2\cdot k_1)].
\label{eq:UUSME}
\end{eqnarray}
Our choice of photon polarization vectors \refeq{eq:poldef} implies
\begin{eqnarray}
\varepsilon_i k_j = 0, \qquad i,j = 1,2 ,
\end{eqnarray}
and thus the SME $\M_{3,5,7,13,15,17}^{\VA,t}$ vanish.
While the SME defined in \refeq{eq:UUSME} are independent in $D$
dimensions, in four dimensions the following relations hold
for the non-vanishing SME
\begin{eqnarray}
0 &=& (t-u) [2\M_1^{{\rm V},t}+\M_2^{{\rm V},t}-2\M_4^{{\rm V},t}
	     -2\M_{11}^{{\rm V},t}]
     + 2s[\M_{11}^{{\rm V},t}-\M_{12}^{{\rm V},t}], \nn\\
0 &=& (ut-\Mt^4)[2\M_1^{{\rm V},t}+\M_2^{{\rm V},t}-2\M_{11}^{{\rm V},t}]
     - 2(t-\Mt^2)(u-\Mt^2)\M_4^{{\rm V},t}
- 2s[\M_{14}^{{\rm V},t}-2\M_{16}^{{\rm V},t}], \nn\\
0 &=& 2\M_1^{{\rm A},t}+\M_2^{{\rm A},t}-2\M_4^{{\rm A},t}.
\label{eq:smerel}
\end {eqnarray}

The $\M_i^{\VA,t}$ of \refeq{eq:UUSME} are defined
such as to obtain
convenient expressions for $t$-channel diagrams. For $u$-channel diagrams it
is useful to introduce a second set $\M_i^{\VA,u}$ of SME
which results from the $\M_i^{\VA,t}$ by the interchange
$(k_1,\veps_1)\leftrightarrow(k_2,\veps_2)$. Of course, the $\M_i^{\VA,u}$
can be completely expressed in terms of the original set $\M_i^{\VA,t}$.

\section{Lowest-order \cs}
\label{se:born}

\begin{figure}
\begin{center}
\begin{picture}(10,3)
\put(-2.7,-14.1){\includegraphics{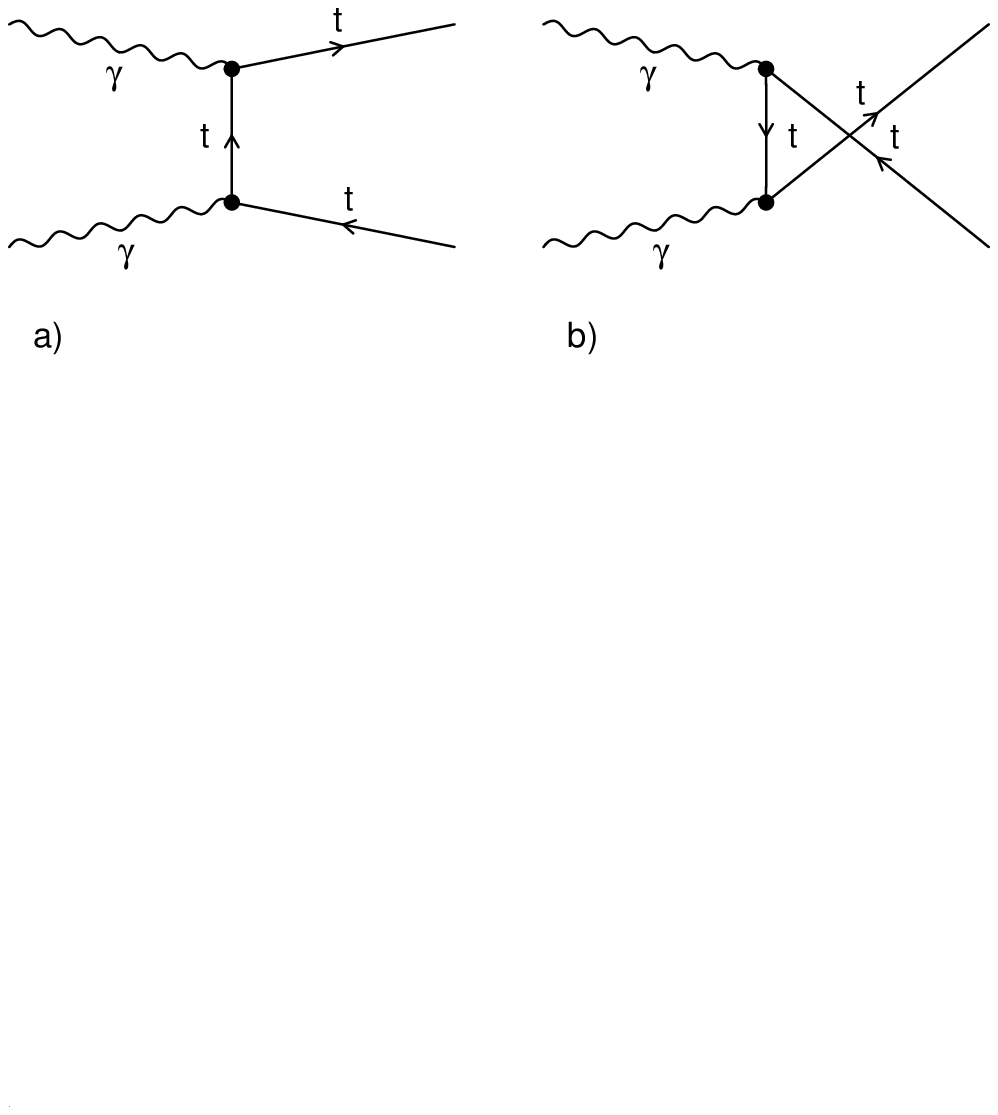}}
\end{picture}
\end{center}
\caption{Lowest-order diagrams for $\AAtt$}
\label{fi:bornab}
\end{figure}

The two tree-level diagrams for $\AAtt$ are shown in \reffi{fi:bornab}.
The corresponding amplitude is given by
\beq
\M_\born = 4\pi Q_\Pt^2\alpha\left(
  \frac{\M_{\born}^t}{t-\Mt^2} +
  \frac{\M_{\born}^u}{u-\Mt^2} \right)
\eeq
with
\beq
\M_{\born}^{\{t,u\}} = \M_{1}^{{\rm V},\{t,u\}} - \M_{11}^{{\rm V},\{t,u\}},
\eeq
leading to the differential Born \cs\
\begin{equation}
\left(\frac{\rd\si_\born}{\rd\Om_\CMS}\right) =
  \sum_{\la_1,\la_2,\si,\bar\si}\frac{1}{4}
  \left(1+P_1\la_1\right)\left(1+P_2\la_2\right)
  \frac{N_\Pt^{\rm C}\be}{64\pi^2s} \left| \M_\born \right|^2,
\label{eq:dsdom}
\end{equation}
where $P_i$ denotes the degree of polarization of the $i$-th photon,
and $N_\Pt^{\rm C}=3$ represents the colour factor for the top quark.

Using the photon polarization states defined in \refeq{eq:poldef}
and summing over the
top-quark polarizations yields
\begin{eqnarray}
\left(\frac{\rd\si_\born^{\rm UU}}{\rd\Om_\CMS}\right) &=&
  Q_\Pt^4\al^2 \frac{3\be}{2s}
  \left[-6\Mt^8+3\Mt^4t^2-\Mt^2t^3+14\Mt^4tu-7\Mt^2t^2u+t^3u\right. \nn\\
&& \left.{}+3\Mt^4u^2-7\Mt^2tu^2-\Mt^2u^3+tu^3\right]/
  \left[(t-\Mt^2)^2(u-\Mt^2)^2\right], \nn\\
\left(\frac{\rd\si_\born^{\pm\pm}}{\rd\Om_\CMS}\right) &=&
  Q_\Pt^4\al^2 \frac{3\be}{s}
  \frac{\Mt^2(s-2\Mt^2)s^2}{(t-\Mt^2)^2(u-\Mt^2)^2}, \nn\\
\left(\frac{\rd\si_\born^{\pm\mp}}{\rd\Om_\CMS}\right) &=&
  Q_\Pt^4\al^2 \frac{3\be}{s}
  \frac{(ut-\Mt^4)(6\Mt^4-4\Mt^2t+t^2-4\Mt^2u+u^2)}{(t-\Mt^2)^2(u-\Mt^2)^2}.
\end{eqnarray}
The superscript ``UU'' denotes to the completely unpolarized case.
Integrating over the symmetric angular range
$\theta_\cut\leq\theta\leq 180^\circ-\theta_\cut$,
we obtain the total \css\
\begin{eqnarray}
\si_\born^{\rm UU} &=&
12\pi Q_\Pt^4\al^2 \frac{1}{s}
  \left[-\tilde{\be}
        -\frac{16\tilde{\be}\Mt^4}{(1-\tilde{\be}^2)s^2}
        +\frac{s^2+4s\Mt^2-8\Mt^4}{s^2}
  \ln\frac{1+\tilde{\be}}{1-\tilde{\be}}\right],
\nn\\
\si_\born^{\pm\pm} &=&
48\pi Q_\Pt^4\al^2 \frac{\Mt^2(s-2\Mt^2)}{s^3}
  \left[\frac{2\tilde{\be}}{1-\tilde{\be}^2}+
  \ln\frac{1+\tilde{\be}}{1-\tilde{\be}}\right], \nn\\
\si_\born^{\pm\mp} &=&
24\pi Q_\Pt^4\al^2 \frac{1}{s}
  \left[-\tilde{\be}
        -\frac{4\tilde{\be}\Mt^2(s+2\Mt^2)}{(1-\tilde{\be}^2)s^2}
        +\frac{s^2+2s\Mt^2-4\Mt^4}{s^2}
  \ln\frac{1+\tilde{\be}}{1-\tilde{\be}}\right],
\end{eqnarray}
where $\tilde{\be}=\be\cos{\theta_\cut}$.

Figures \ref{fi:bornw0} and \ref{fi:bornw20} show the \css\
integrated over
two different angular ranges ($\theta_\cut=0^\circ,20^\circ$)
for various polarization configurations.
The angular cut reduces in particular
$\si^{\pm\pm}_\born$ at high energies
considerably.
For energies close to threshold the \css\ for
unequal photon helicities are suppressed, more precisely
$\si_\born^{\pm\mp}$
vanish like
$\be^3$ for $\be\to 0$ while $\si_{\born}^{\pm\pm}$
behave like $\be$.
In the high-energy limit (with fixed $\theta_\cut$) the
\css\ $\si^{\pm\mp}_\born$ drop like
$\ln(s)/s$ and $1/s$ for $\theta_\cut=0$ and $\theta_\cut\ne 0$,
respectively,
whereas $\si_\born^{\pm\pm}$ behave like $1/s$ and $1/s^2$ without and
with finite angular cut-off, respectively.

The angular dependence of the differential \css\ is illustrated
in \reffi{fi:difcs} at $\sqrt{s} = 350$, $500$,
$1000\GeV$.
The \css\ $\rd\si_\born^{\pm\pm}/{\rd\Om_\CMS}$
are maximal in the forward and backward directions owing to the
approximative $t$- and $u$-channel poles. These poles are cancelled in
$\rd\si_\born^{\pm\mp}/{\rd\Om_\CMS}$ by a kinematical zero
such that these \css\ vanish in the forward and backward directions.
For low energies they possess a
maximum at $90^\circ$ which is split into two maxima for
CMS energies above
$\sqrt{2(1+\sqrt5)}\Mt\approx 2.54\Mt$.
For higher energies
these maxima occur at $\theta[{\rm rad}] \sim \Mt/E$ and $\pi-\Mt/E$ so that
all channels become more and more peaked in
the forward and backward directions.

\section{Radiative corrections}
\label{se:RC}

\subsection{Structure of the $\Oa$ corrections and Feynman diagrams}

Using standard techniques (e.g.\ described in \citere{ADHab}), we have
calculated the complete electroweak virtual and soft-photonic
corrections to $\AAtt$ in $\Oa$ for arbitrary photon polarizations.
We have performed two completely independent calculations which agree
numerically within 9--10 digits for the considered energies.
The $\Oa$-corrected differential \cs\ takes the general form
\begin{eqnarray}
\left(\frac{\rd\si}{\rd\Om_\CMS}\right) &=&
  \sum_{\la_1,\la_2,\si,\bar\si}
  \frac{1}{4}\left(1+P_1\la_1\right)\left(1+P_2\la_2\right)
  \frac{N_\Pt^{\rm C}\be}{64\pi^2s}
  \left[\left|\M_\born\right|^2
  (1+\de_{\rm SB})+2\Re\left\{\M^\ast_\born\de\M
  \right\}\right]
\nn\\[.3em]
&=& \left(\frac{\rd\si_\born}{\rd\Om_\CMS}\right)(1+\de),
\label{eq:radcor}
\end{eqnarray}
where $\de_{\rm SB}$ denotes the soft-photon bremsstrahlung factor in
$\Oa$, and $\de\M$ the one-loop corrections to the transition amplitude.
The factor $\de$ represents the complete relative $\Oa$ correction.
For the integrated \cs\ $\si$, it is defined analogously
\beq
\si = \int_{\theta_{\min}}^{\theta_{\max}} \rd\!\cos\theta \int_0^{2\pi}
\rd\phi\,\left(\frac{\rd\si}{\rd\Omega}\right) = \si_{\Born}(1+\delta).
\eeq

In the soft-photon approximation for the bremsstrahlung process
$\ga\ga\to\Pt\Ptbar\ga$ only photons
with energies below the cut-off energy $\De E\ll E$
are included.
The IR divergence is regulated by an infinitesimal photon mass $\la$; it
cancels
against the IR divergence of the virtual corrections.
The soft-photonic correction leads to the following
correction factor to the lowest-order \cs\
\beqar
\delta_{\SB} &=& -\frac{\al}{\pi}
  \biggl\{2\ln\frac{2\De E}{\la}
  +\frac{1}{\be}\ln\biggl(\frac{1-\be}{1+\be}\biggr)
  +\frac{s-2\Mt^2}{s\be}\biggl[2\ln\frac{2\De E}{\la}
         \ln\biggl(\frac{1-\be}{1+\be}\biggr)
\nn \\ \phantom{-\frac{\al}{\pi}\biggl\{}
&& {}-2\Li_2\biggl(\frac{1-\be}{1+\be}\biggr)+\frac{1}{2}
       \ln^2\biggl(\frac{1-\be}{1+\be}\biggr)+\frac{\pi^2}{3}
     -2\ln\biggl(\frac{1-\be}{1+\be}\biggr)
      \ln\biggl(\frac{2\be}{1+\be}\biggr)\biggr]\biggr\},
\hspace{2em}
\eeqar
which can be easily obtained from the general results of \citere{ADHab}.

The one-loop Feynman diagrams, which form the virtual $\Oa$ correction
$\de\M$, have been calculated in 't Hooft--Feynman gauge.
The calculation of $\de\M$ can be split into two different steps.
Firstly, one has to evaluate
the interference of the SME with the Born matrix elements
for fixed photon polarizations
\begin{equation} \label{eq:contractions}
M^{\la_1\la_2,\{V,A\},\{t,u\}}(i,j) =
  \sum_{\sigma,\bar\sigma} \M_{i}^{\{V,A\},\{t,u\}} \times
  \M_{\rm Born}^{j \: \ast},
\end{equation}
with $i = 1,\ldots,7,11,\ldots,17$ and $j = t,u$.
These quantities are listed in \refapp{se:ConSME}. The corresponding
expressions for unpolarized photons, which can be obtained by averaging over
the photon polarizations, were already given in \citere{ggtt}.
Note, however, that we use a different definition of the quantities
$\M_{\rm Born}^{\{t,u\}}$.
Secondly, one has to calculate the one-loop contributions
$\de F^{\rm \{V,A\}}_i$ to the
invariant functions $F^{\rm \{V,A\}}_i$ defined in \refeq{eq:sum1}.
Technically the $\de F^{\rm \{V,A\}}_i$ are expressed in terms of
one-loop tensor integrals.
The amplitude is generated with {\it FeynArts}
\cite{fa}. Its reduction to
coefficients of one-loop tensor integrals is
carried out in {\it Mathematica} \cite{math}, once using {\it FeynCalc}
\cite{fc} once without.
Following the method of \citere{pas79},
the tensor coefficients are recursively calculated from the scalar
one-loop integrals, which are evaluated according to \citere{scalar}.
We have checked our numerical routines for the calculation of the scalar
one-loop integrals against the package {\it FF}
\cite{FF}.

The renormalization has been carried out in the complete on-shell scheme
\cite{ADHab}, where all mass parameters represent physical masses, and
all physical fields possess residues equal to one. Consequently,
self-energy insertions in external legs of Feynman graphs do not
contribute. We have checked that all UV singularities cancel.

The relevant one-loop diagrams are classified with respect
to their topology in self-energy, vertex and box corrections. Since the
analytical form of the complete virtual corrections is very complicated
and untransparent, we do not give the full explicit expressions. Instead,
we list all relevant one-loop Feynman graphs and discuss only the most
important radiative corrections analytically.

\begin{figure}
\begin{center}
\begin{picture}(16,6)
\put(-2.6,-12.0){\includegraphics{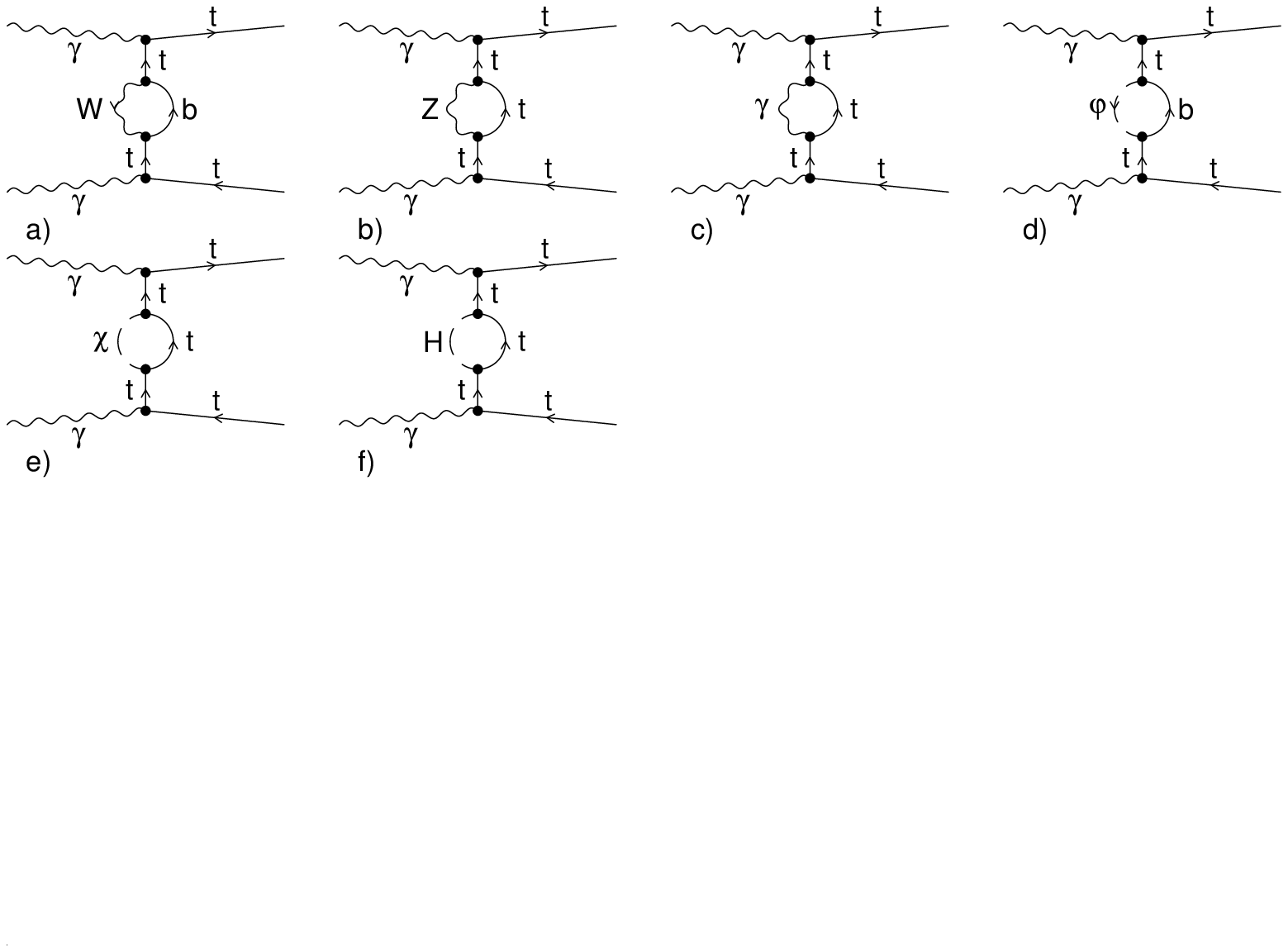}}
\put(9.5,1.0){+ \quad crossed graphs}
\end{picture}
\caption{Top-quark self-energy diagrams
\label{fi:sediag} }
\end{center}
\begin{center}
\begin{picture}(16,8.5)
\put(-2.7,-8.9){\includegraphics{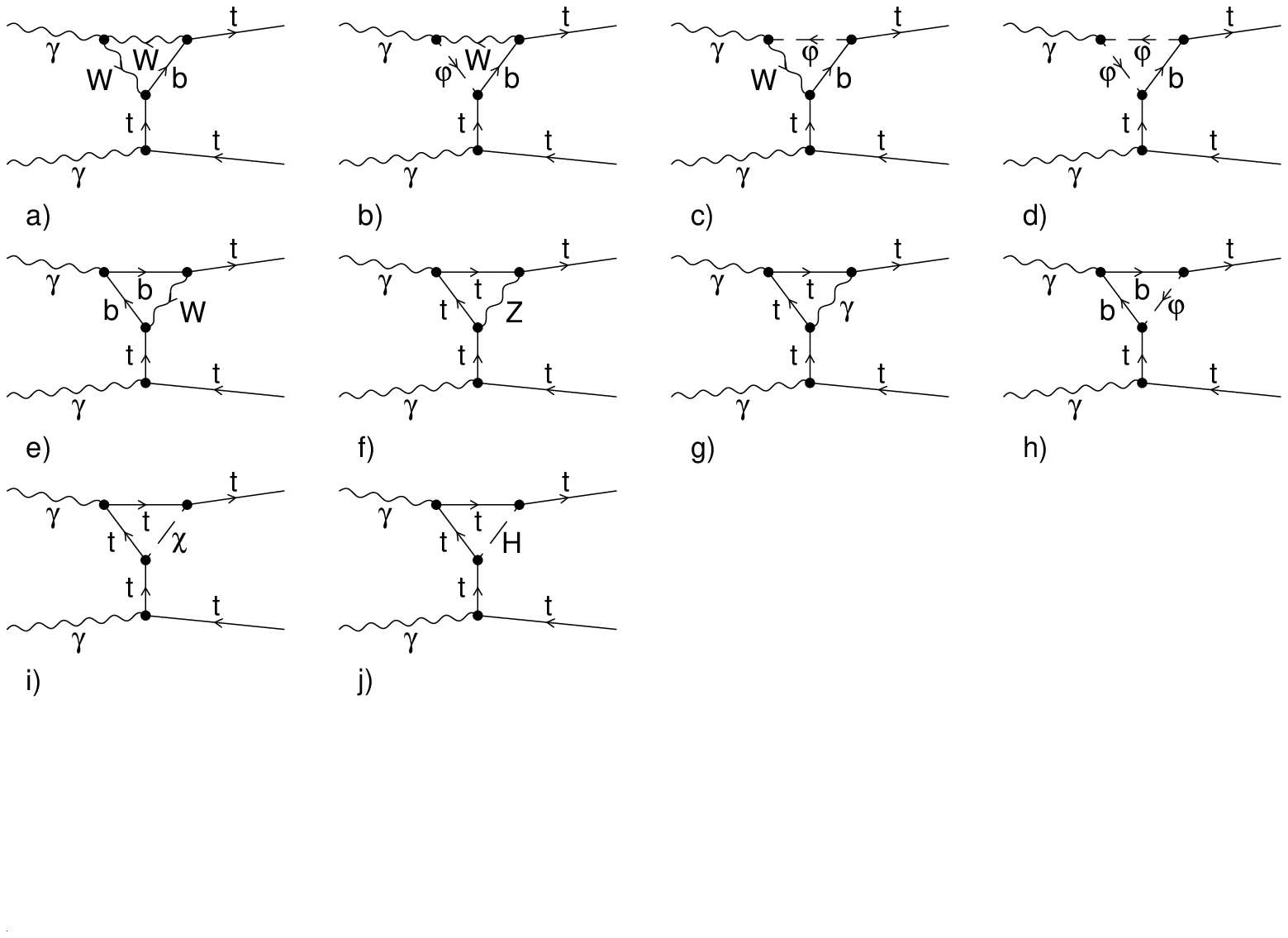}}
\put(9.5,1.5){+ \quad crossed graphs}
\end{picture}
\end{center}
\caption{$At^\ast\bar{t}$ vertex diagrams
\label{fi:Attdiag} }
\end{figure}

The top-quark self-energy is the only two-point function occurring at one
loop. The corresponding $t$-channel diagrams are shown in \reffi{fi:sediag}.
The $u$-channel diagrams are obtained from those
via ``crossing'', i.e.\ by interchanging the incoming photon lines.
For renormalization only the $ZA$-mixing energy is needed in addition.

\begin{figure}
\begin{center}
\begin{picture}(12.5,3)
\put(-2.6,-14.7){\includegraphics{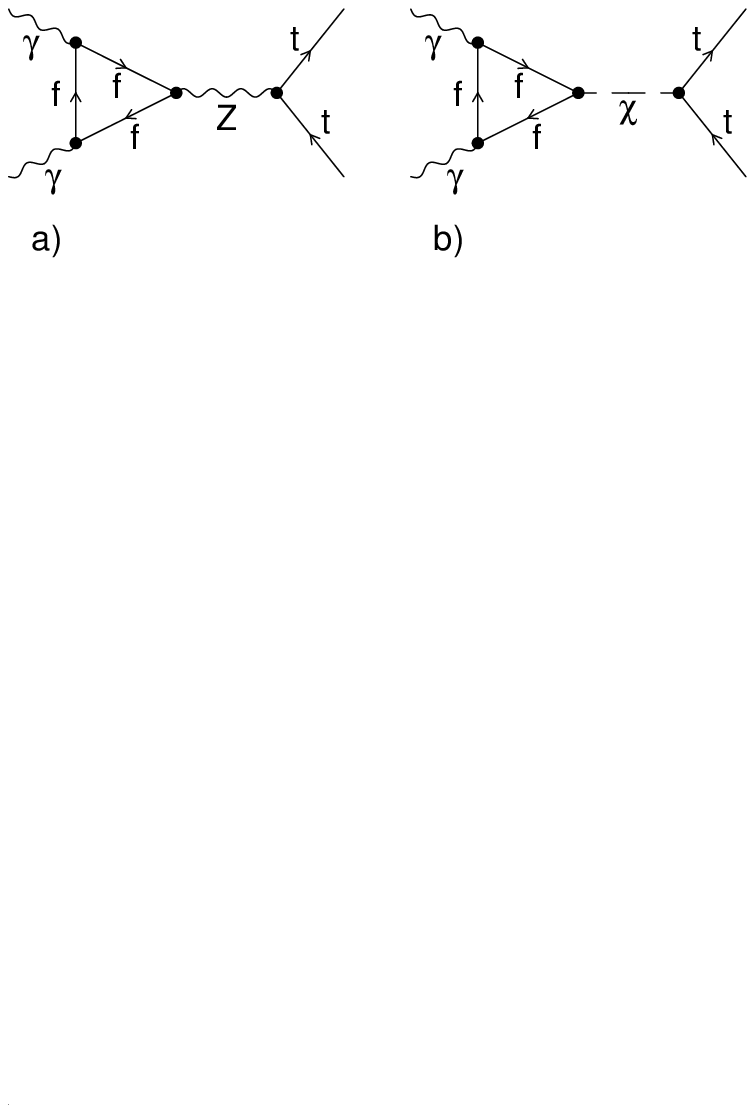}}
\put(9.0,1.2){+ \quad crossed graphs}
\end{picture}
\end{center}
\caption{$AAZ^{\ast}$ and $AA\chi^{\ast}$ vertex corrections
\label{fi:AAZdiag} }
\begin{center}
\begin{picture}(16,14)
\put(-2.6, -3.2){\includegraphics{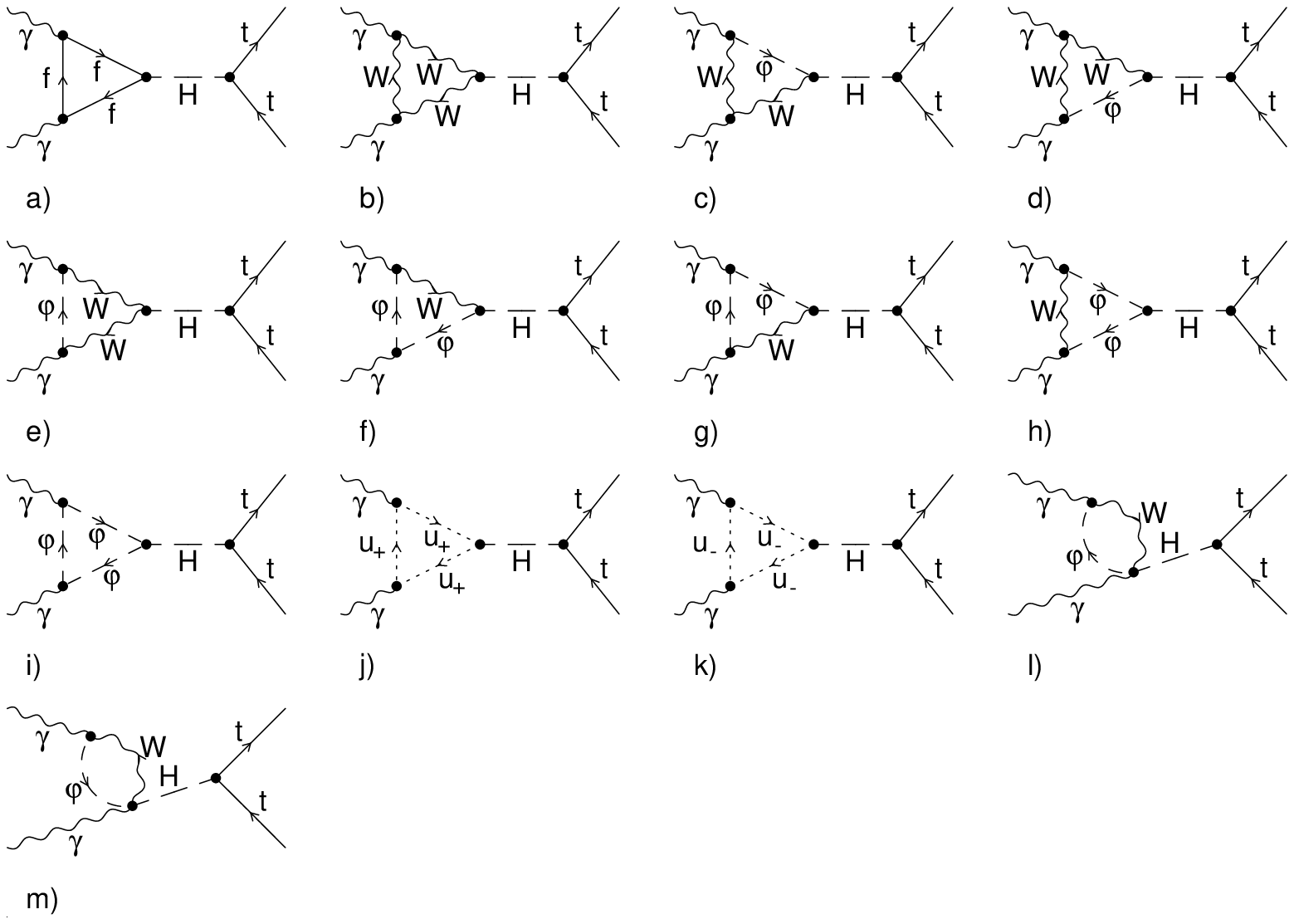}}
\put(-2.6,-14.8){\includegraphics{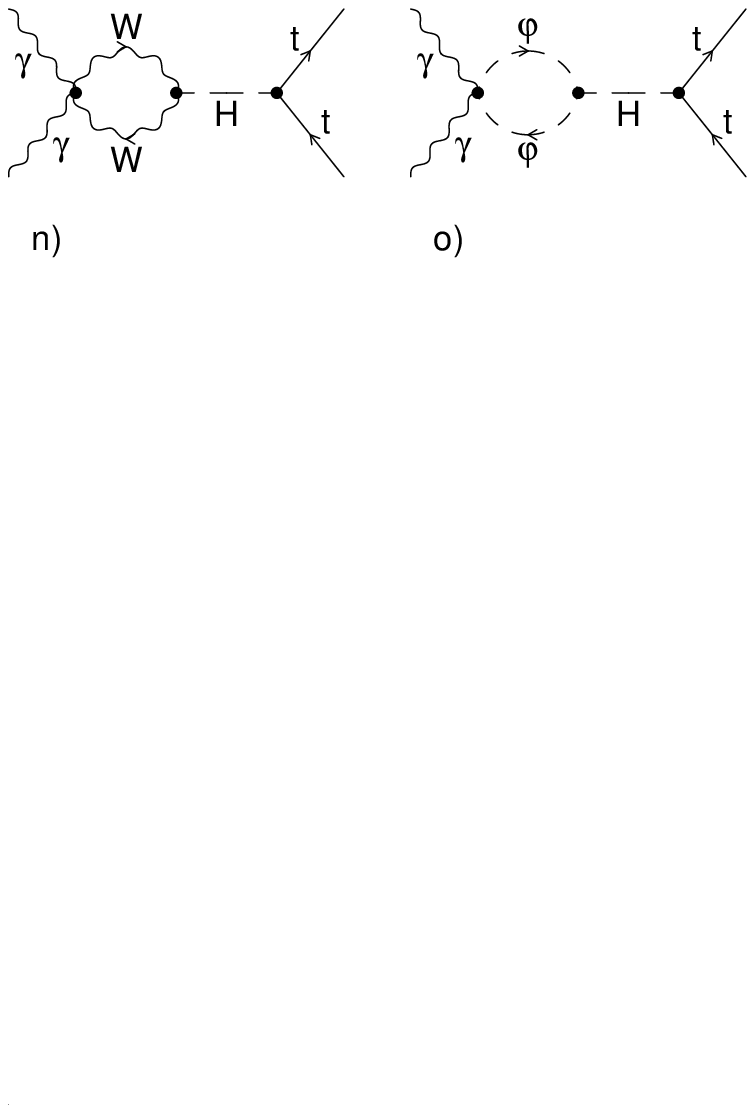}}
\put(6.0,4.3){+ \quad crossed graphs}
\end{picture}
\end{center}
\caption{$AAH^{\ast}$ vertex corrections (Graphs n,o do not possess
crossed counterparts.)
\label{fi:AAHdiag} }
\end{figure}

The vertex corrections can be divided into three very different classes.
Firstly, there are $At^\ast\bar{t}$ and $At\bar{t}^\ast$
vertex corrections in the $t$ and $u$ channel, where asterisks mark
off-shell fields. The Feynman graphs for the
$At^\ast\bar{t}$ vertex corrections are shown in \reffi{fi:Attdiag},
the $At\bar{t}^\ast$ vertex graphs are obtained from
those by
reversing the top-quark lines.
The second and third type of vertex corrections concern the
$AAZ^\ast/\chi^\ast$ and
$AAH^\ast$ couplings, respectively, with $\chi$ being the neutral
would-be Goldstone boson field related to the \PZ~boson.
The ($s$-channel-like)
diagrams are depicted in \reffis{fi:AAZdiag} and \ref{fi:AAHdiag}, and
the corresponding corrections are discussed in more detail in the
subsequent sections.

\begin{figure}
\begin{center}
\begin{picture}(16,17)
\put(-2.6, -3.2){\includegraphics{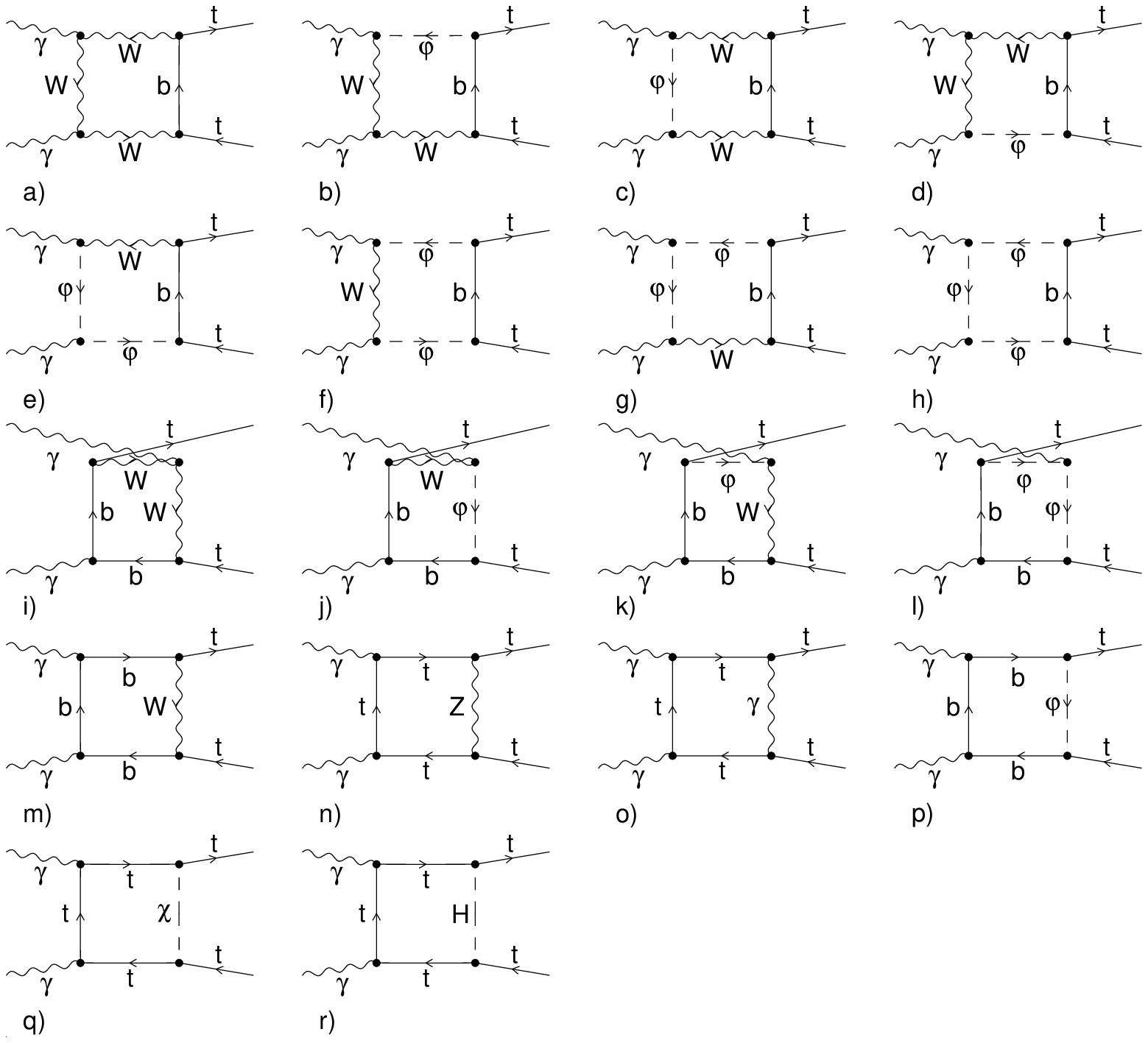}}
\put(-2.6,-14.8){\includegraphics{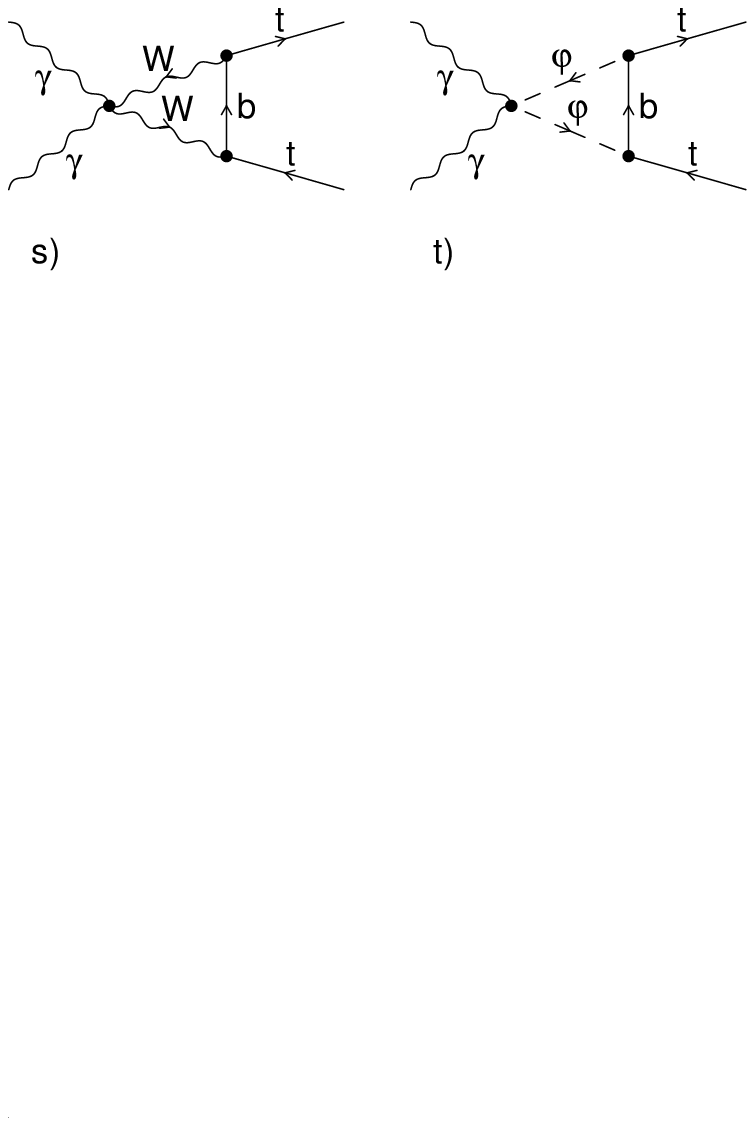}}
\put(9.0,4.3){+ \quad crossed graphs}
\end{picture}
\end{center}
\caption{Box corrections (Graphs s,t do not possess
crossed counterparts.)
\label{fi:boxdiag} }
\end{figure}

The irreducible one-loop diagrams, so-called boxes, are shown in
\reffi{fi:boxdiag}.

\subsection{Gauge-invariant subsets of Feynman diagrams}

The classification of Feynman graphs according to their topology is
very convenient for practical calculations, but as a matter of fact not
based on physical grounds. This is due to the fact that the single vertex
functions in general do not represent gauge-invariant subsets of diagrams.
On the other hand, it is very desirable to decompose the complicated
structure of the complete $\Oa$ corrections into parts which are of
different physical origin and can be discussed separately. Indeed, the
fact that we omit here the QCD corrections is already such a separation.

The process $\AAtt$ can also be treated in pure QED, which is part of the
SM. Consequently, all QED-like diagrams, i.e.\ the ones with only $A$ and $t$
fields as virtual lines, form a gauge-invariant subset. Thus, we define
the one-loop QED corrections $\de_\QED$ by the sum of the soft-photon
correction
$\de_\SB$ and the contribution of the photonic one-loop diagrams shown in
\ref{fi:sediag}.c, \ref{fi:Attdiag}.g and \ref{fi:boxdiag}.o, their
counterparts arising from particle interchange, and the corresponding
terms in the top-quark field and mass renormalization.

Since the number of fermion generations in the SM is a free
parameter of the theory, each fermion generation yields a gauge-invariant
subset of one-loop diagrams formed by fermionic loops. For $\AAtt$ this
set is furnished by the diagrams \ref{fi:AAZdiag}.a,
\ref{fi:AAZdiag}.b,
\ref{fi:AAHdiag}.a, and their crossed counterparts. In the following we
call these fermion-loop corrections $\de_\ferm$.
Actually, the fermion-loop corrections to the $AAH^\ast$ vertex on the
one hand and to the $AAZ^\ast/\chi^\ast$ vertices on the other hand are
separately gauge-invariant. This becomes clear e.g.\ by inspecting the
process $\AAtt$ within the gauged non-linear $\si$-model \cite{gnlsm}.
This model represents an SU(2)$\times$U(1) gauge theory either, which is
equivalent to the SM in the unitary gauge without physical Higgs boson.
However, in this model without Higgs field the $AAZ^\ast/\chi^\ast$
corrections are identical with the ones in the SM.

Of course, all virtual diagrams which are not contained in $\de_\QED$ or
$\de_\ferm$ form a gauge-invariant subset, too. We call the corresponding
corrections ``bosonic'' and denote it by $\de_\bos$. In summary,
we have the decomposition of the $\Oa$ corrections $\de$
\beq
\de = \de_\QED+\de_\ferm+\de_\bos.
\eeq

In the following we
mainly concentrate on the non-QED or ``weak''
corrections defined as
\beq
\de_\weak = \de_\ferm+\de_\bos.
\eeq

\subsection{\AAZ$^{\ast}$ and \AAX$^{\ast}$ vertex corrections}
\label{se:fermcor}

In 't Hooft--Feynman gauge the fermion-loop contributions to the
$AAZ^\ast$ and $AA\chi^\ast$ vertices (see \reffi{fi:AAZdiag}) are given
by
\beqar
\de\M^{AAZ^{\ast}} &=& \frac{2\al^2}{\sw^2\cw^2}
  \frac{s}{(u-t)(s-\MZ^2)}
[\M_{11}^{{\rm V},t}-\M_{12}^{{\rm V},t}]
\sum_f N_f^{\rm C} Q_f^2 I_{W,f}^3 m_f^2 C(s,m_f),
\nn\\
\de\M^{AA\chi^{\ast}} &=& -\frac{s}{\MZ^2}\de\M^{AAZ^{\ast}},
\label{eq:AAZcor}
\eeqar
where the relations \refeq{eq:smerel} between the SME have been used,
and $\cw=\MW/\MZ$, $\sw=\sqrt{1-\cw^2}$ denote the sine and cosine of the
weak mixing angle.
The sum over $\sum_f$ runs over all (massive) fermions $f$ with colour
factor $N_f^{\rm C}$, electric
charge $Q_f$, third component of weak isospin $ I_{W,f}^3$, and mass $m_f$.
In \refeq{eq:AAZcor} $C(s,m)$ denotes the scalar three-point function
(see e.g.\ \citere{ADHab})
\beq
C(s,m) = C_0(s,0,0,m,m,m) =
\frac{1}{2s}\ln^2\left(\frac{r+1}{r-1}\right), \qquad \mbox{with} \quad
r = \sqrt{1-\frac{4m^2}{s+i\eps}}.
\label{eq:C0mmm}
\eeq
Note that the pole at $s=\MZ^2$ drops out in the gauge-invariant sum of
$\de\M^{AAZ^{\ast}}$ and $\de\M^{AA\chi^{\ast}}$,
\beqar
\de\M^{AAZ^{\ast}+AA\chi^{\ast}} &=&
\de\M^{AAZ^{\ast}}+\de\M^{AA\chi^{\ast}}
\nn\\ &=&
\frac{2\al^2}{\sw^2\MW^2}
  \frac{s}{(t-u)}
[\M_{11}^{{\rm V},t}-\M_{12}^{{\rm V},t}]
\sum_f N_f^{\rm C} Q_f^2 I_{W,f}^3 m_f^2 C(s,m_f) \nl
&=& - \frac{2\al^2}{\sw^2\MW^2}
   \la_1 \sqrt{s}\,\Mt\,\de_{\la_1\la_2}\,\de_{\si\bar\si}
\sum_f N_f^{\rm C} Q_f^2 I_{W,f}^3 m_f^2 C(s,m_f).
\label{eq:AAZcor2}
\eeqar
Since $\de\M^{AAZ^{\ast}+AA\chi^{\ast}}$ is gauge-invariant, it is equal to
$\de\M^{AAZ^{\ast}}$ in the unitary gauge, which has been checked
explicitly.

As can be seen in \refeq{eq:AAZcor2},
this correction only contributes for equally polarized photons.
The top-loop contribution is enhanced by a factor $\Mt^2/(\sw^2\MW^2)$.
Close to threshold an additional enhancement results from the
scalar 3-point function which tends to
\beq
C(s,m_f) \sim -\frac{\pi^2}{2s}
\eeq
such that these diagrams give rise to corrections of about $10\%$.

\subsection{Higgs resonance}
\label{se:Higres}

The contribution of the $AAH^{\ast}$ vertex corrections
(see \reffi{fi:AAHdiag}) reads (compare \citere{Ve94})
\beq
\de\M^{AAH^{\ast}} = \frac{F^H(s)}{s-\MH^2} \M_{12}^{{\rm V},t}
= -\frac{F^H(s)}{s-\MH^2}\,\be\,\sqrt{s}\,\Mt\,{\rm sgn}(\si)
   \,\de_{\la_1\la_2}\,\de_{\si\bar\si},
\label{eq:Hres}
\eeq
where
\beqar
F^H(s) &=& \frac{\al^2}{2\sw^2}
  \Biggl\{ {}-2 \sum_f N_f^{\rm C}Q_f^2\frac{\Mf^2}{\MW^2}
  \left[ 2+(4\Mf^2-s) \, C(s,m_f) \right]
\nn\\ &&
\phantom{\frac{\al^2}{2\sw^2}\frac{\M_{12}^{{\rm V},t}}{s-\MH^2}\biggl\{}
  {}+\left[ \frac{\MH^2}{\MW^2}+6+(\MH^2+12\MW^2-7s) \, C(s,\MW) \right]
  \Biggr\},
\label{eq:FH}
\eeqar
with $C(s,m)$ defined in \refeq{eq:C0mmm}.
{}From \refeq{eq:Hres} we recognize that $\de\M^{AAH^{\ast}}$ vanishes
unless the photons
are equally polarized.
While the fermionic part of $F^H(s)$ is gauge-invariant, the
bosonic part depends on the gauge. Equation
\refeq{eq:FH} is written down in
't~Hooft--Feynman gauge. Up to a proportionality factor
containing the SME, $\de\M^{AAH^{\ast}}$ has the same analytical structure
as the corresponding corrections to $\AAWW$ \cite{AAWW} evaluated
in 't~Hooft--Feynman gauge.

For $s\to\MH^2$ the correction $\de\M^{AAH^{\ast}}$ becomes resonant, i.e.\
finite-width effects of the Higgs boson have to be taken into account.
Introducing
a constant or energy-dependent width via the substitution
$(s-\MH^2)^{-1}\to(s-\MH^2+i\MH\Ga_\PH)^{-1}$ clearly breaks gauge
invariance. Instead, we split $\de\M^{AAH^{\ast}}$ into a gauge-invariant
resonant and gauge-dependent non-resonant part, and introduce a constant
width only in the former,
\begin{equation}
\frac{F^H(s)}{s-\MH^2} \;\;\to\;\;
  \frac{F^H(\MH^2)}{s-\MH^2+i\MH\Gamma_\PH}+
  \frac{F^H(s)-F^H(\MH^2)}{s-\MH^2}.
\end{equation}
For a calculation with order $\Oa$ accuracy also near $s=\MH^2$,
one should take into account the $\Oa$ corrections to the
Higgs-boson width \cite{Fl81} and to
$F^H(s)$ in the resonant
contribution, i.e.\ for $s\sim\MH^2$. Since the
Higgs resonance is not our main concern, we only take into account the
lowest-order decay width determined from the imaginary part of the
one-loop Higgs-boson \se\ and \refeq{eq:FH} for $F^H(\MH^2)$.

\subsection{Special features of the electroweak one-loop corrections}

Here we discuss some interesting peculiarities of the electroweak
one-loop corrections to $\AAtt$.
These are related to the fact that this process involves no light
charged external particles and that the tree-level matrix elements
involve only QED couplings. Similar features have been observed in the
process $\AAWW$ \cite{AAWW}.
\renewcommand{\labelenumi}{(\roman{enumi})}
\begin{enumerate}
\item Coulomb singularity \\
Near threshold every pair-production process of charged particles gets
large photonic corrections which are due to the well-known {\it Coulomb
singularity}. The relative correction behaves like $\be^{-1}$, more
precisely it approaches asymptotically the universal factor
\beq
\de\si_\Coul = \frac{\al\pi}{2\be} \, Q_\Pt^2 \, \si_\Born.
\label{eq:coul}
\eeq
Diagrammatically this correction originates from the box
diagram \ref{fi:boxdiag}.o and its crossed counterpart and is
contained in the QED part $\de_\QED$ of the complete $\Oa$ correction.
The modifications of the Coulomb singularity owing to the finite
top-quark width have been discussed in \citere{Fa87}.

\item Leading logarithmic QED corrections \\
Because $\AAtt$ involves no light charged external particles, no large
logarithmic corrections associated with collinear photons arise for not
too high energies. Consequently, the QED corrections are of the same
order as the weak corrections.

\item Heavy Higgs effects \\
In the limit $\MH\gg s,\Mt$, for fixed $\al$, $\MZ$, $\MW$,
no large logarithmic corrections of the form
$\ln(\MH^2/|q^2|)$ with $q^2=s,t,u,\Mt^2$ occur. In other words the
$\Oa$ correction $\de$
approaches a constant in the formal limit $\MH\to\infty$. The absence of
$\ln\MH$ terms is associated with the fact that the one-loop correction
to $\AAtt$ is UV-finite in the gauged non-linear $\si$-model despite of
its non-renormalizability. Inspecting also the finite terms in the
heavy-Higgs limit, one finds that the SM one-loop corrections for
$\MH\to\infty$ exactly coincide with the ones of the gauged non-linear
$\si$-model. This has been explicitly shown in \citere{HHSM}, where the
physical Higgs field of the SM was integrated out.

\item Running $\al(q^2)$ and $\rho$-parameter \\
Coupling constant renormalization often leads to large universal
corrections entering via the running of the electromagnetic coupling
$\al(q^2)$ or the $\rho$-parameter. For $\AAtt$ no such corrections exist
since on-shell photons naturally couple with $\al(0)=\al$, i.e.\ completely
independent of the weak mixing angle.
Technically, the cancellation of the electromagnetic vacuum-polarization
effects between charge and photonic field renormalization is due to a
Ward identity.

\end{enumerate}

\section{Numerical results}
\label{se:numres}

For the numerical evaluation we
have
used the following set of parameters
\cite{PDG94}
\beq
\begin{array}[b]{lcllcllcl}
\alpha &=& 1/137.0359895, &
\GF & = & 1.166390 \times 10^{-5} \GeV^{-2}, \\[.3em]
\MZ & = & 91.187\GeV, &
\MH & = & 250\GeV, &&& \\[.3em]
\Me & = & 0.51099906\MeV,  \hspace{1.5em} &
m_{\mu} & = & 105.65839\MeV,  \hspace{1.5em} &
m_{\tau} & = & 1.777\;\GeV, \\[.3em]
\Mu & = & 46.0\;\MeV, &
\Mc & = & 1.50\;\GeV, &
\Mt & = & 170\;\GeV, \\[.3em]
\Md & = & 46.0\;\MeV, &
\Ms & = & 150\;\MeV, &
\Mb & = & 4.50\;\GeV.
\end{array}
\label{eq:par}
\eeq
The masses of the light quarks are adjusted such that the
experimentally measured hadronic vacuum polarization is reproduced
\cite{Ei95}.
As the Fermi-constant $\GF$ is empirically much better
known than the \PW~mass, \MW\ is calculated from all the other
parameters using the muon decay width including radiative corrections.
In this calculation of \MW\ all parameters given above enter sensibly.
If not stated otherwise, $\MW$ is determined in the following using
formulae (2.56) and (2.57) of \citere{wwrev}. The above set of
parameters yields
$$\MW=80.333\GeV.$$

\begin{figure}
\setlength{\unitlength}{1mm}
\begin{picture}(160,210)(-1,0)
\put(120,175){\begin{picture}(60,30)
               \setlength{\unitlength}{1pt}
               \put(0,30){\begin{tabular}{rl} UU : & \solid  \\
                             $\pm\pm$ : & \dashed \\ $\pm\mp$ : & \dashdotted
			  \end{tabular}}
               \end{picture}}
\put(10,117){\makebox(130,82)
          {\includegraphics{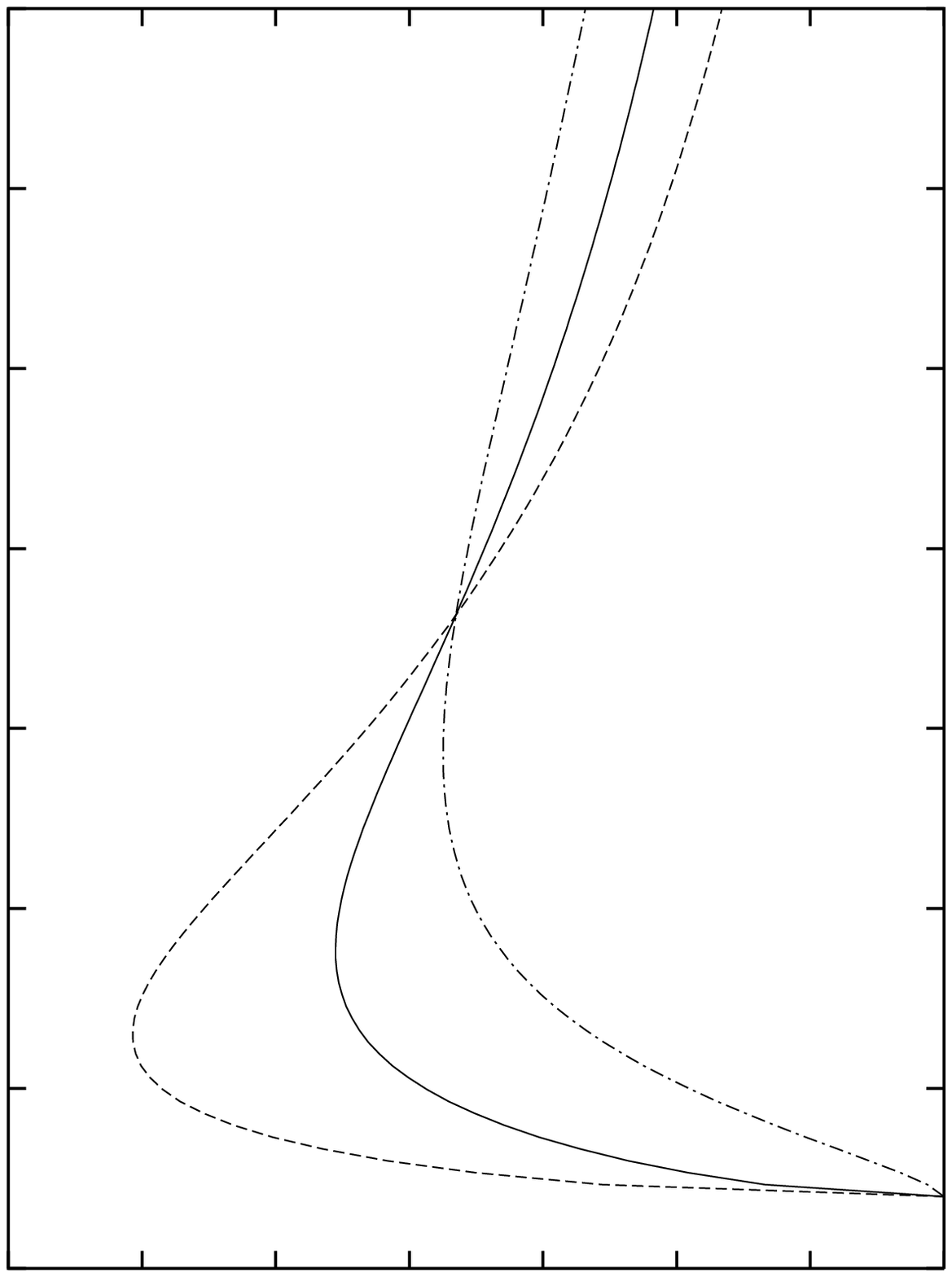}}}
\put(10,117){\begin{picture}(130,80)(0,0)
               \setlength{\unitlength}{1mm}
               \put(17,-5){$400$ }
               \put(58,-5){$600$ }
               \put(99.5,-5){$800$ }
               \put(139,-5){$1000$ }
               \put(128.5,-10){$E_{\rm CMS}/$GeV}
               \put(-10,87){$\sigma_{\rm Born}/$pb}
               \put(-4.5,-1.5){$0$}
               \put(-7,22.5){$0.4$}
               \put(-7,46.5){$0.8$}
               \put(-7,70.5){$1.2$}
               \end{picture}}
\put(10,12){\makebox(130,82)
          {\includegraphics{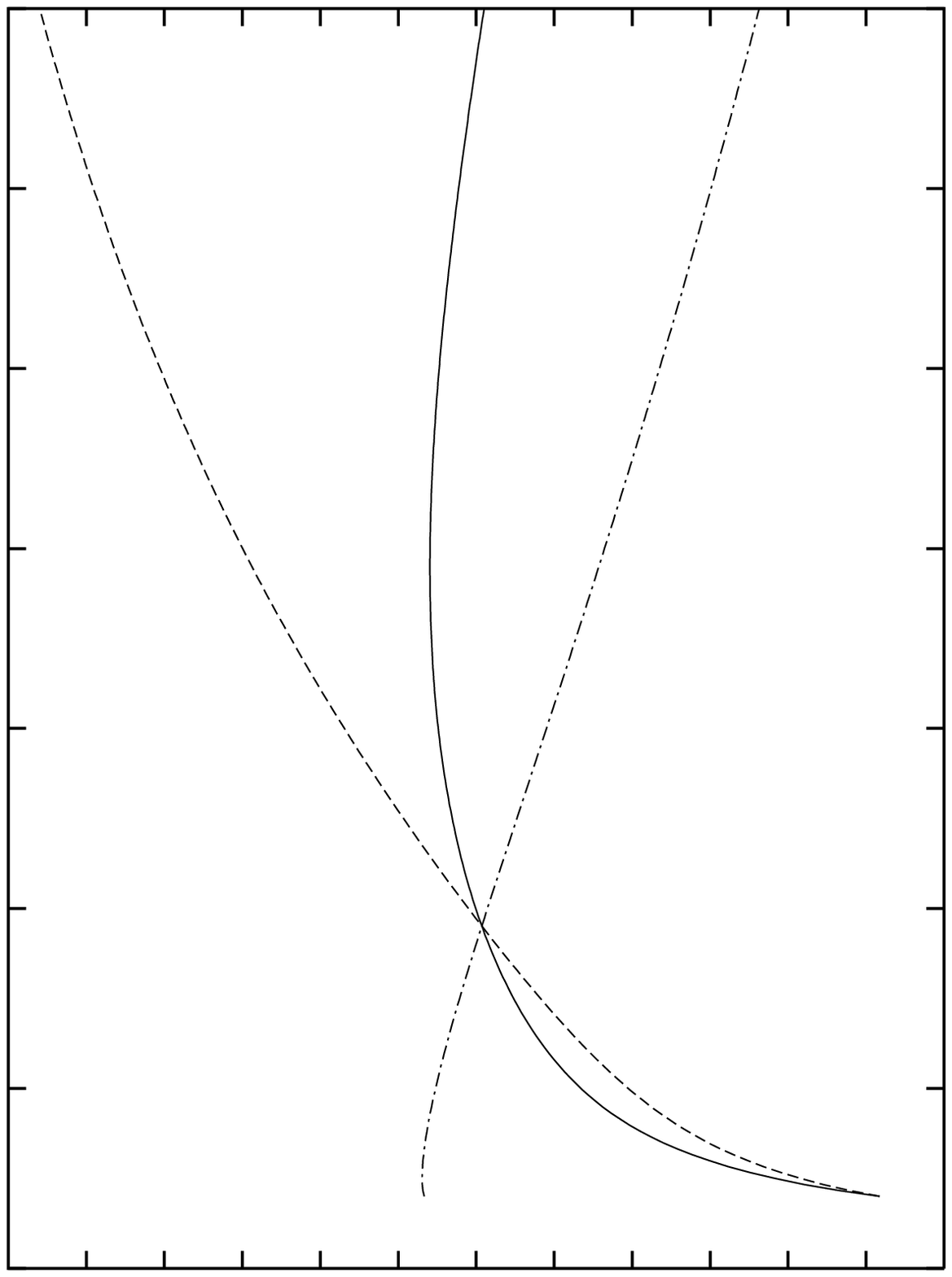}}}
\put(10,12){\begin{picture}(130,80)(0,0)
               \setlength{\unitlength}{1mm}
               \put(17,-5){$400$ }
               \put(58,-5){$600$ }
               \put(99.5,-5){$800$ }
               \put(139,-5){$1000$ }
               \put(128.5,-10){$E_{\rm CMS}/$GeV}
               \put(-7,88){$\delta_{\weak}$/\%}
               \put(-9.5,6){$-10$ }
               \put(-7.8,20.5){$-8$ }
               \put(-7.8,34){$-6$ }
               \put(-7.8,48){$-4$ }
               \put(-7.8,61){$-2$ }
               \put(-4.5,74.5){$0$ }
               \end{picture}}
\end{picture}
\caption{Integrated lowest-order cross-sections and corresponding relative
  corrections for several polarizations in the full angular range}
\label{fi:bornw0}
\end{figure}

Figure \ref{fi:bornw0} shows the \cs\ integrated
over the full angular range and the corresponding weak corrections for
different polarizations of the incoming photons.
In the energy range from threshold up to $1\TeV$ the weak
corrections range between 0 and $-10\%$.
For oppositely polarized photons the corrections are about $-4\%$ close to
threshold and increase smoothly to $-8\%$ at $1\TeV$.
For equally polarized photons the corrections amount to roughly $-10\%$
close to threshold,
tend to positive values with increasing energy and reach zero at about
$1\TeV$.
The large negative corrections close to threshold originate from
top-quark loop corrections to the $AAZ^\ast/\chi^\ast$ vertices that are
enhanced by the large top-quark mass (see \refse{se:fermcor}).
\begin{figure}
\setlength{\unitlength}{1mm}
\begin{picture}(160,210)(-1,0)
\put(120,175){\begin{picture}(60,30)
               \setlength{\unitlength}{1pt}
               \put(0,30){\begin{tabular}{rl} UU : & \solid  \\
                             $\pm\pm$ : & \dashed \\ $\pm\mp$ : & \dashdotted
			  \end{tabular}}
               \end{picture}}
\put(10,117){\makebox(130,82)
          {\includegraphics{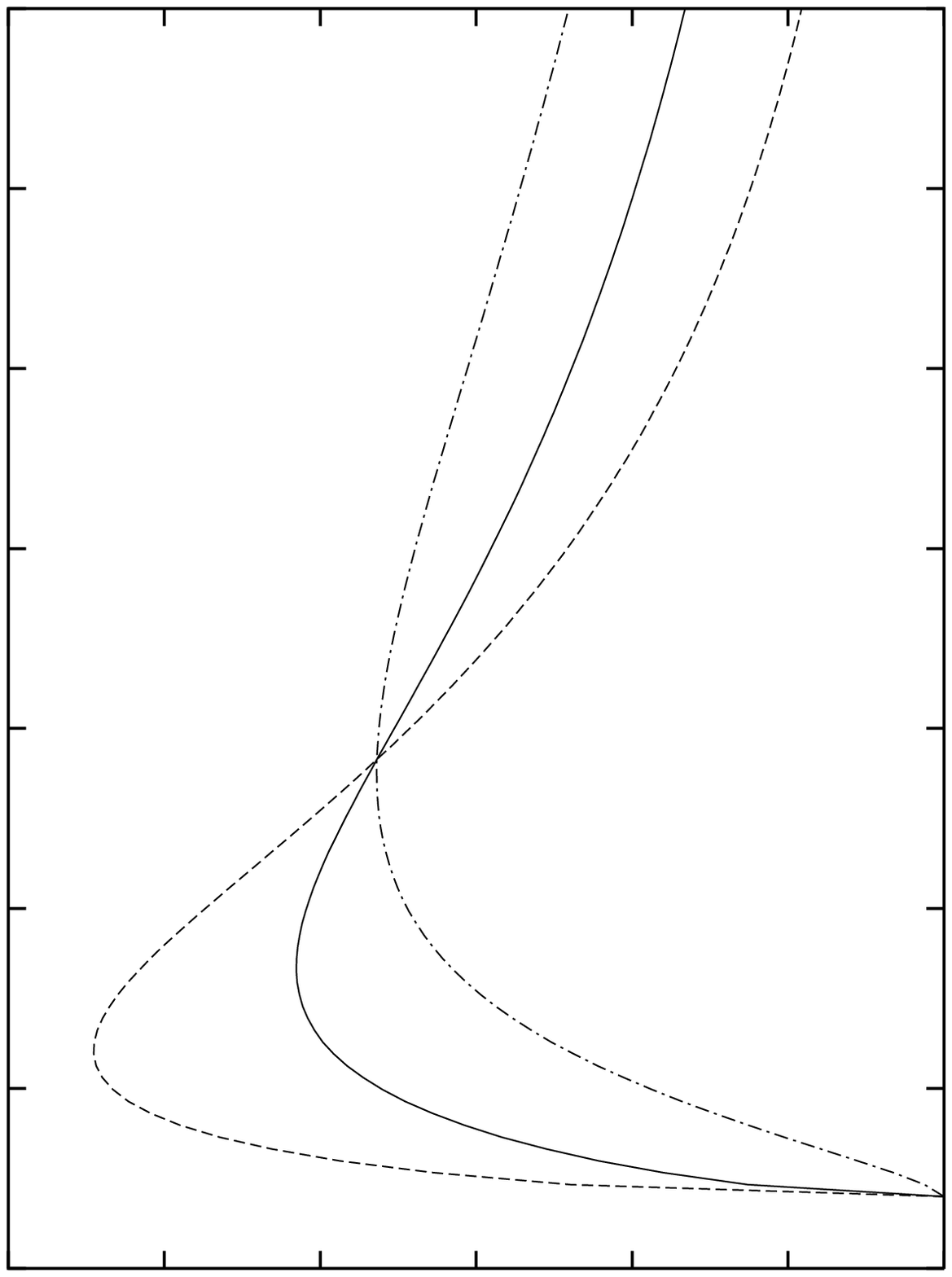}}}
\put(10,117){\begin{picture}(130,80)(0,0)
               \setlength{\unitlength}{1mm}
               \put(17,-5){$400$ }
               \put(58,-5){$600$ }
               \put(99.5,-5){$800$ }
               \put(139,-5){$1000$ }
               \put(-10,87){$\sigma_{\rm Born}/$pb}
               \put(128.5,-10){$E_{\rm CMS}/$GeV}
               \put(-4.5,-1.5){$0$}
               \put(-7,26.5){$0.4$}
               \put(-7,54.5){$0.8$}
               \put(-7,81.5){$1.2$}
               \end{picture}}
 \put(10,12){\makebox(130,82)
          {\includegraphics{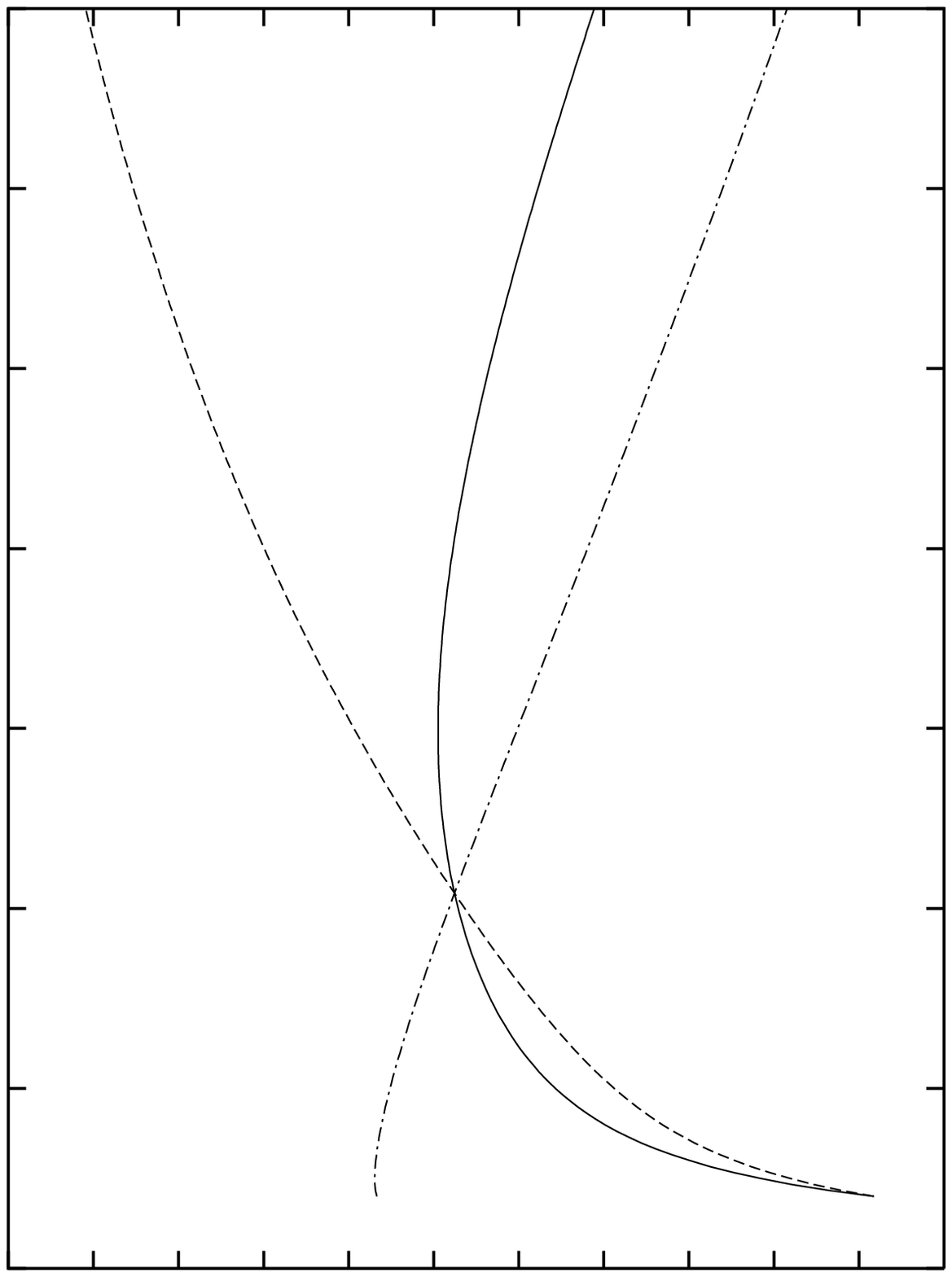}}}
\put(10,12){\begin{picture}(130,80)(0,0)
               \setlength{\unitlength}{1mm}
               \put(17,-5){$400$ }
               \put(58,-5){$600$ }
               \put(99.5,-5){$800$ }
               \put(139,-5){$1000$ }
               \put(128.5,-10){$E_{\rm CMS}/$GeV}
               \put(-7,88){$\delta_{\weak}$/\% }
               \put(-9.5,6.5){$-10$ }
               \put(-7.8,21.5){$-8$ }
               \put(-7.8,36.5){$-6$ }
               \put(-7.8,51.8){$-4$ }
               \put(-7.8,67){$-2$ }
               \put(-4.5,81.5){$0$ }
               \end{picture}}
\end{picture}
\caption{Same as in \reffi{fi:bornw0} but with an angular cut
         $20^\circ \leq \theta \leq 160^\circ$}
\label{fi:bornw20}
\end{figure}
Integrating only over the restricted angular range
 $20^\circ \leq \theta \leq 160^\circ$ affects the corrections only
weakly, as can be seen from \reffi{fi:bornw20}.

\begin{figure}
\setlength{\unitlength}{1mm}
\begin{picture}(160,210)(-1,0)
\put(13,101){\makebox(0,0)[l]{{ $E_{\rm CMS} = 350$ GeV}}}
\put(65,101){\makebox(0,0)[l]{{ $E_{\rm CMS} = 500$ GeV}}}
\put(117,101){\makebox(0,0)[l]{{ $E_{\rm CMS} = 1000$ GeV}}}
\put(57,202){\begin{picture}(60,30)
               \setlength{\unitlength}{1pt}
               \put(0,15){UU : $\rule[.85mm]{11.8mm}{.2mm}$,}
               \put(0,0){$\pm \pm$ : $\rule[.85mm]{2.2mm}{.2mm} \hspace{1mm}
                 \rule[.85mm]{2.2mm}{.2mm} \hspace{1mm}
                 \rule[.85mm]{2.2mm}{.2mm} \hspace{1mm}
                 \rule[.85mm]{2.2mm}{.2mm} \hspace{1mm}$, }
               \put(75,15){$+-$ : $\rule[.85mm]{2.2mm}{.2mm} \hspace{1mm}
                 \rule[.85mm]{.3mm}{.3mm} \hspace{1mm}
                 \rule[.85mm]{2.2mm}{.2mm} \hspace{1mm}
                 \rule[.85mm]{.3mm}{.3mm}$ }
               \put(75,0){$-+$ : $\rule[.85mm]{2mm}{.2mm} \hspace{1mm}
                 \rule[.85mm]{2mm}{.2mm} \hspace{3mm}
                 \rule[.85mm]{2mm}{.2mm} \hspace{1mm}
                 \rule[.85mm]{2mm}{.2mm}$}
               \end{picture}}

\put(10,117) {\makebox(40,80)
          {\includegraphics{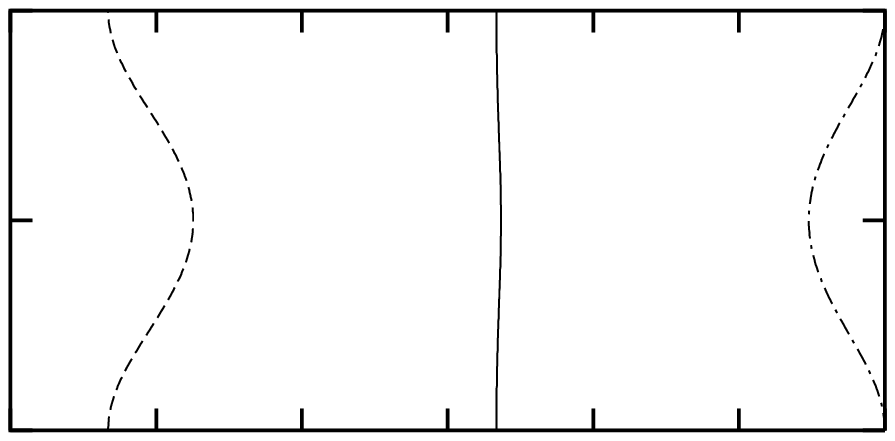}}}
\put(10,117){\begin{picture}(40,80)(0,0)
               \setlength{\unitlength}{1mm}
               \put(-2,-5){$0^{\circ}$ }
               \put(18,-5){$90^{\circ}$ }
               \put(37,-5){$180^{\circ}$ }
               \put(44,-10){$\theta$ }
               \put(-10,86){$ \left( \frac{{\textstyle \sigma}_{\rm Born}}
                {{\textstyle \Omega}_{\rm CMS}} \right) / $ pb}
               \put(-4.5,-1.5){$0$}
               \put(-8.5,25.5){$0.02$}
               \put(-8.5,52.5){$0.04$}
               \put(-8.5,79){$0.06$}
               \end{picture}}
\put(62,117) {\makebox(40,80)
          {\includegraphics{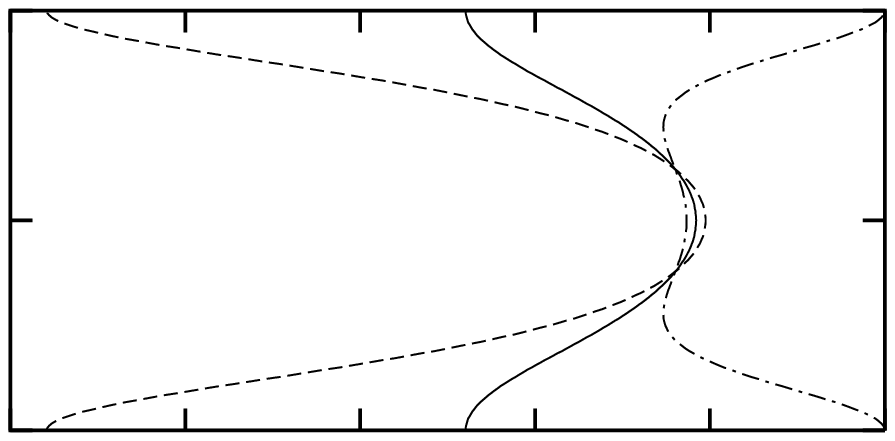}}}
\put(62,117){\begin{picture}(40,80)(0,0)
               \setlength{\unitlength}{1mm}
               \put(-2,-5){$0^{\circ}$ }
               \put(18,-5){$90^{\circ}$ }
               \put(37,-5){$180^{\circ}$ }
               \put(44,-10){$\theta$ }
               \put(-4.5,-1.5){$0$}
               \put(-7,30.5){$0.1$}
               \put(-7,62.5){$0.2$}
               \end{picture}}
\put(114,117) {\makebox(40,80)
          {\includegraphics{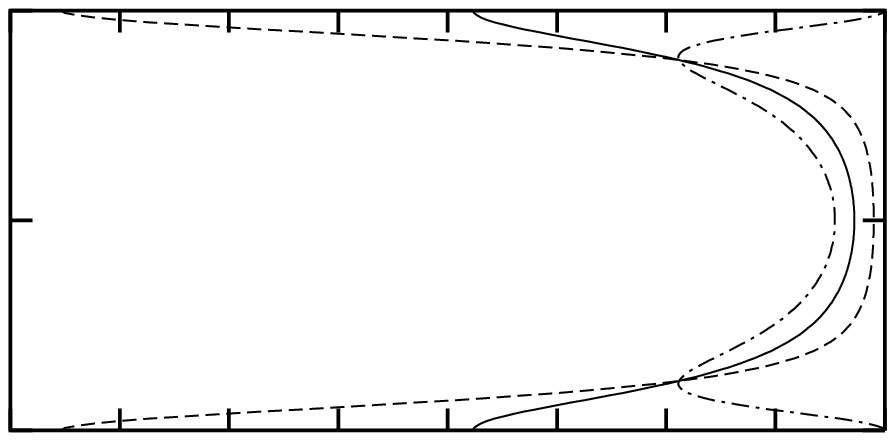}}}
\put(114,117){\begin{picture}(40,80)(0,0)
               \setlength{\unitlength}{1mm}
               \put(-2,-5){$0^{\circ}$ }
               \put(18,-5){$90^{\circ}$ }
               \put(37,-5){$180^{\circ}$ }
               \put(44,-10){$\theta$ }
               \put(-4.5,-1.5){$0$}
               \put(-7,18.5){$0.1$}
               \put(-7,38.5){$0.2$}
               \put(-7,58.5){$0.3$}
               \put(-7,78.5){$0.4$}
               \end{picture}}
\put(10,10) {\makebox(40,80)
          {\includegraphics{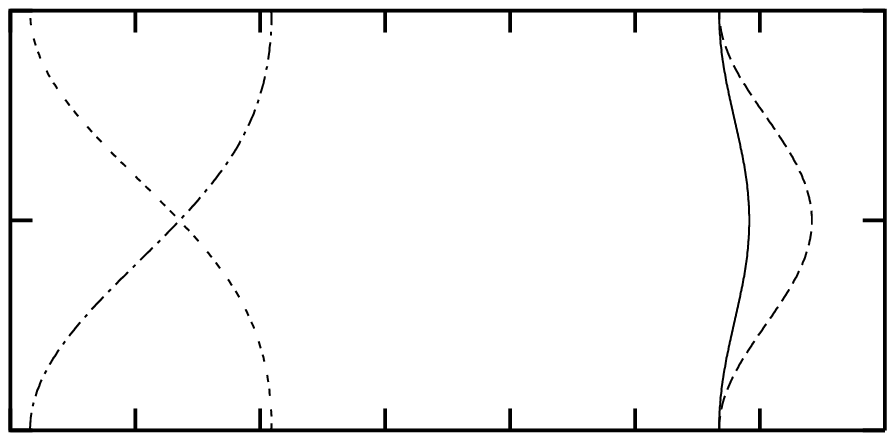}}}
\put(10,10){\begin{picture}(40,80)(0,0)
               \setlength{\unitlength}{1mm}
               \put(-2,-5){$0^{\circ}$ }
               \put(18,-5){$90^{\circ}$ }
               \put(37,-5){$180^{\circ}$ }
               \put(44,-10){$\theta$ }
               \put(-9.5,-1.5){$-10$}
               \put(-7.8,21.5){$-8$}
               \put(-7.8,44.5){$-6$}
               \put(-7.8,67.5){$-4$}
               \put(-5,84){$\delta_{\rm weak}$ / \% }
               \end{picture}}
\put(62,10) {\makebox(40,80)
          {\includegraphics{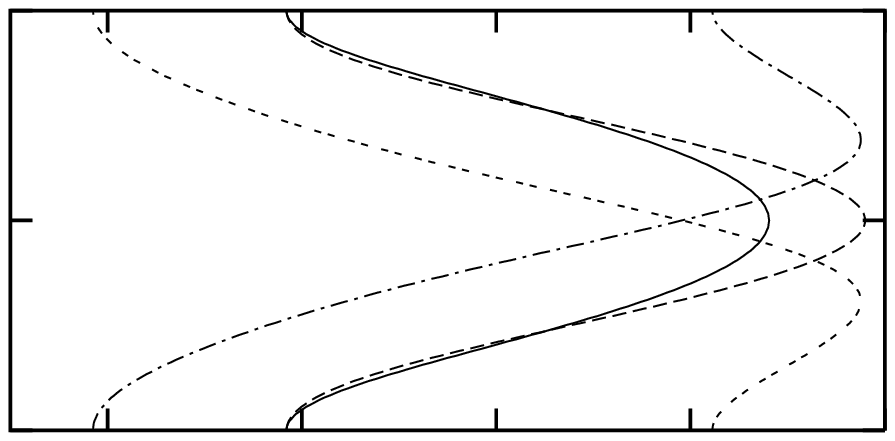}}}
\put(62,10){\begin{picture}(40,80)(0,0)
               \setlength{\unitlength}{1mm}
               \put(-2,-5){$0^{\circ}$ }
               \put(18,-5){$90^{\circ}$ }
               \put(37,-5){$180^{\circ}$ }
               \put(44,-10){$\theta$ }
               \put(-7.8,-1.5){$-8$}
               \put(-7.8,16.5){$-6$}
               \put(-7.8,34.5){$-4$}
               \put(-7.8,52){$-2$}
               \put(-4.5,69.5){$0$}
               \end{picture}}
\put(114,10) {\makebox(40,80)
          {\includegraphics{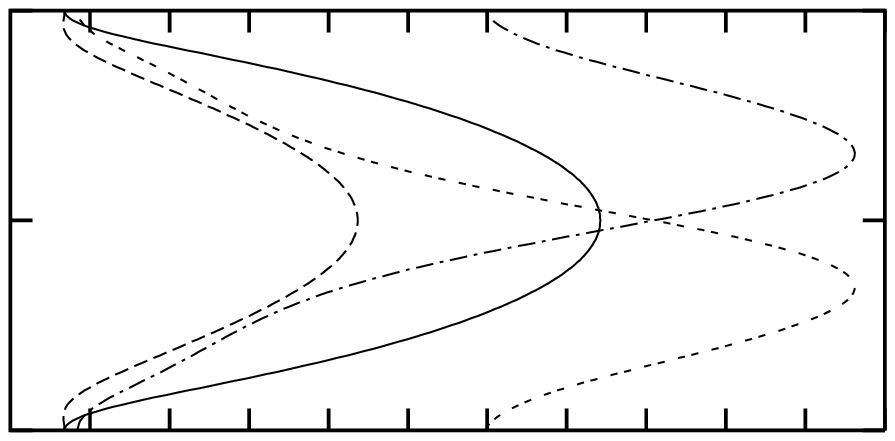}}}
\put(114,10){\begin{picture}(40,80)(0,0)
               \setlength{\unitlength}{1mm}
               \put(-2,-5){$0^{\circ}$ }
               \put(18,-5){$90^{\circ}$ }
               \put(37,-5){$180^{\circ}$ }
               \put(44,-10){$\theta$ }
               \put(-9.5,-1.5){$-18$}
               \put(-9.5,21){$-12$}
               \put(-7.8,42.5){$-6$}
               \put(-4.5,64){$0$}
               \end{picture}}
\end{picture}
  \caption{Differential lowest-order \css\ and relative
   corrections for the unpolarized \cs\ and the \css\ with equal  and
   unequal photon helicities}
\label{fi:difcs}
\end{figure}

In \reffi{fi:difcs} we plot the weak corrections to the differential
\css\ for $\sqrt{s}=350$, $500$,
$1000\GeV$.
The angular dependence of the corrections increases with energy.
For equally polarized or unpolarized
photons the corrections are forward--backward symmetric and take their
maximum for $\theta=90^\circ$. They are usually negative but
become
positive
at high energies in the forward and backward
directions.
In the case of oppositely polarized photons the corrections are asymmetric with
a maximum that moves from the forward or backward direction towards
$\theta=90^\circ$ with increasing energy.
This maximum reaches almost $-18\%$ at $1\TeV$.

In \refta{ta:res1UU} we list numerical values for the unpolarized \cs\ and the
corresponding corrections for several energies and scattering angles.
\begin{table}
\arraycolsep 5pt
\newdimen\digitwidth
\setbox0=\hbox{0}
\digitwidth=\wd0
\catcode`!=\active
\def!{\kern\digitwidth}
\newdimen\minuswidth
\setbox0=\hbox{$-$}
\minuswidth=\wd0
\catcode`?=\active
\def?{\kern\minuswidth}
\begin{center}
$$
\begin{array}{|c|c||c|c|c|c|c|c|}
\hline
E_\CMS & \theta & \si_\Born^{\rm UU}/\mbox{pb} &
\de^{\rm UU}/\mbox{\%} &
\de_\QED^{\rm UU}/\% & \rule[-3mm]{0mm}{8mm}
\de_\weak^{\rm UU}/ \% &
\de_\ferm^{\rm UU}/ \% &
\de_\bos ^{\rm UU}/ \% \\
\hline \hline
& 20^\circ & 0.02663 & !-6.74 & ?1.96 & !-8.70 & -7.86 & !-0.84 \\ \cline{2-8}
& 45^\circ & 0.02653 & !-6.84 & ?1.95 & !-8.79 & -7.72 & !-1.07 \\ \cline{2-8}
350\GeV
& 90^\circ & 0.02634 & !-6.97 & ?1.94 & !-8.92 & -7.55 & !-1.36 \\ \cline{2-8}
& 20^\circ \leq \theta \leq 160^\circ
           & 0.31230 & !-6.89 & ?1.95 & !-8.84 & -7.65 & !-1.19 \\ \cline{2-8}
& !0^\circ \leq \theta \leq 180^\circ
           & 0.33248 & !-6.88 & ?1.95 & !-8.83 & -7.67 & !-1.17 \\
\hline\hline
& 20^\circ & 0.10877 & !-2.40 & ?0.21 & !-2.61 & -2.97 & ?!0.36 \\ \cline{2-8}
& 45^\circ & 0.07929 & !-4.53 & ?0.18 & !-4.71 & -2.92 & !-1.79 \\ \cline{2-8}
500\GeV
& 90^\circ & 0.05395 & !-6.66 & ?0.14 & !-6.81 & -3.14 & !-3.67 \\ \cline{2-8}
& 20^\circ \leq \theta \leq 160^\circ
           & 0.81785 & !-5.12 & ?0.17 & !-5.29 & -3.01 & !-2.28 \\ \cline{2-8}
& !0^\circ \leq \theta \leq 180^\circ
           & 0.90439 & !-4.82 & ?0.17 & !-4.99 & -3.01 & !-1.99 \\
\hline\hline
& 20^\circ & 0.09971 & !-1.75 & -0.46 & !-1.29 & ?0.02 & !-1.31 \\ \cline{2-8}
& 45^\circ & 0.03367 & !-7.66 & -0.57 & !-7.09 & ?0.02 & !-7.11 \\ \cline{2-8}
1\TeV
& 90^\circ & 0.01392 & -11.48 & -0.64 & -10.84 & ?0.03 & -10.87 \\ \cline{2-8}
& 20^\circ \leq \theta \leq 160^\circ
           & 0.33208 & !-7.45 & -0.56 & !-6.89 & ?0.03 & !-6.91 \\ \cline{2-8}
& !0^\circ \leq \theta \leq 180^\circ
           & 0.43447 & !-5.63 & -0.53 & !-5.10 & ?0.03 & !-5.13 \\
\hline
\end{array}
$$
\caption{Lowest-order \css\ and relative corrections for unpolarized photons
($\de_\QED$ is evaluated at $\De E = 0.1E$)}
\label{ta:res1UU}
\end{center}
\end{table}
We include the complete corrections corresponding to a soft-photon energy
cut-off $\De E/E=0.1$,
the QED corrections, the weak corrections and the
fermionic and bosonic
ones. These numbers show that the fermionic
corrections dominate the unpolarized \cs\ at low energies.
At high energies the fermionic corrections are very small.
They  are completely absent for oppositly polarized photons. However,
in this case the bosonic corrections are large (compare the figures).
On the other hand,
the bosonic corrections are usually below 1--2\% for
equally polarized photons.

\begin{figure}
\setlength{\unitlength}{1mm}
\begin{picture}(160,98)(-1,4)
\put(100,75){\begin{picture}(60,30)
              \setlength{\unitlength}{1pt}
               \put(0,15){\begin{tabular}{rl} $\MH=350\GeV$ : & \solid  \\
                              $\MH=400\GeV$ : & \dashed \\
			      $\MH=500\GeV$ : & \dashdotted
			  \end{tabular}}
              \end{picture}}
 \put(10,12){\makebox(130,82)
          {\includegraphics{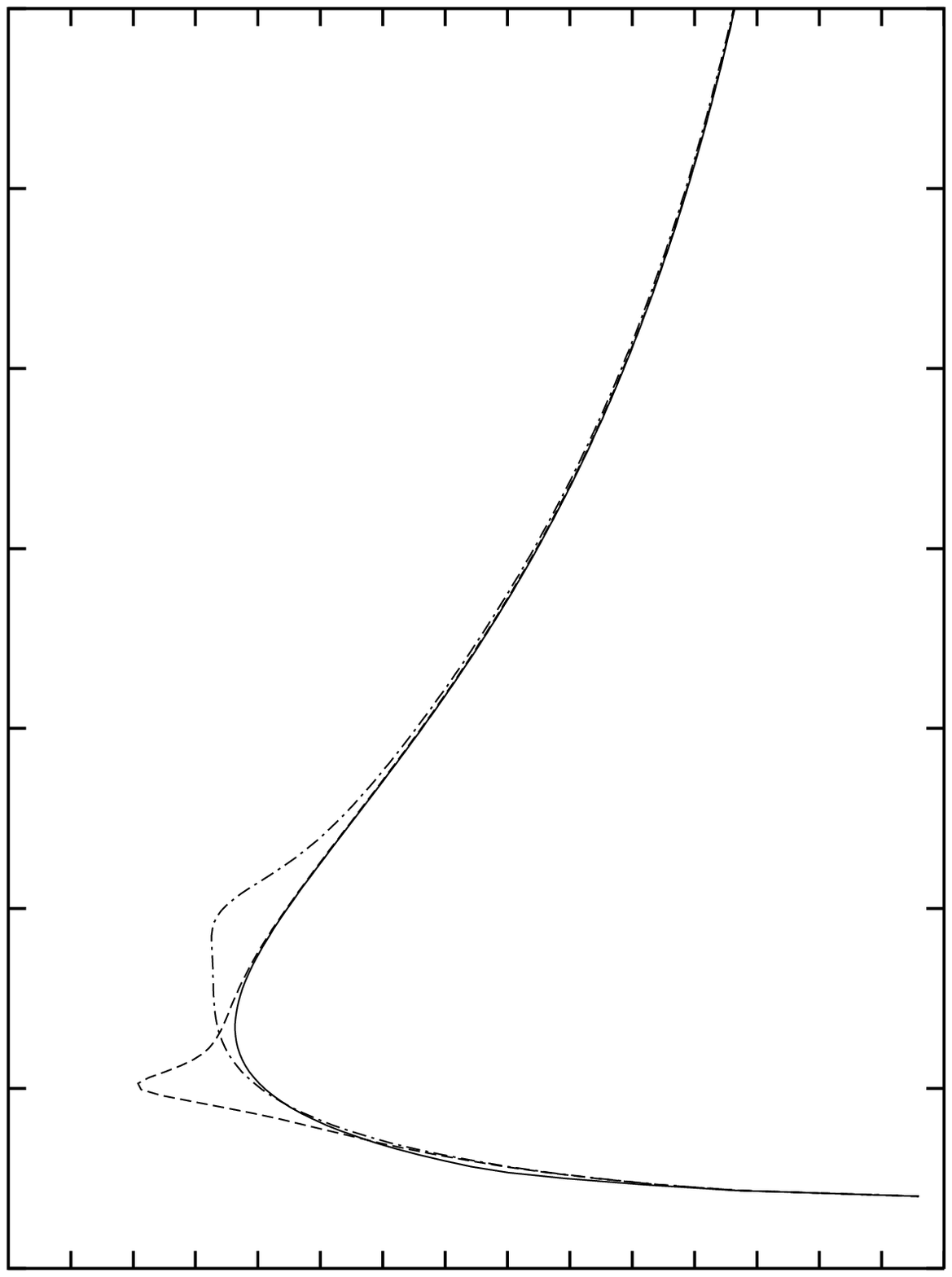}}}
\put(10,12){\begin{picture}(130,80)(0,0)
               \setlength{\unitlength}{1mm}
               \put(17,-5){$400$ }
               \put(58,-5){$600$ }
               \put(99.5,-5){$800$ }
               \put(139,-5){$1000$ }
               \put(128.5,-10){$E_{\rm CMS}/$GeV}
               \put(-5,88){$\sigma^{\pm\pm}$/pb }
               \put(-4.5,-1.5){$0$}
               \put(-7,26.5){$0.5$}
               \put(-4.5,54){$1$}
               \put(-7,82){$1.5$}
               \end{picture}}
\end{picture}
\caption{Integrated \cs\ ($0^\circ<\theta<180^\circ$) for equally
polarized photons including one-loop corrections for various
Higgs masses}
\label{fi:Higgsres}
\end{figure}

In \reffi{fi:Higgsres} we present the \cs\ for equally polarized photons
including the weak corrections integrated over
$0^\circ<\theta<100^\circ$ for the Higgs-boson masses
$\MH=350$, $400$, $500\GeV$.
Owing to the relatively high threshold for $\AAtt$ and the large width
of such a heavy Higgs boson, the effect of the Higgs resonance is
comparably small. In fact a resonance structure is only visible in the
range $400\GeV\lsim\MH\lsim500\GeV$.

\begin{figure}
\setlength{\unitlength}{1mm}
\begin{picture}(160,100)(-1,4)
\put(120,50){\begin{picture}(60,30)
               \setlength{\unitlength}{1pt}
               \put(0,30){\begin{tabular}{rl} UU : & \solid  \\
                             $\pm\pm$ : & \dashed \\ $\pm\mp$ : & \dashdotted
			  \end{tabular}}
               \end{picture}}
 \put(10,12){\makebox(130,82)
          {\includegraphics{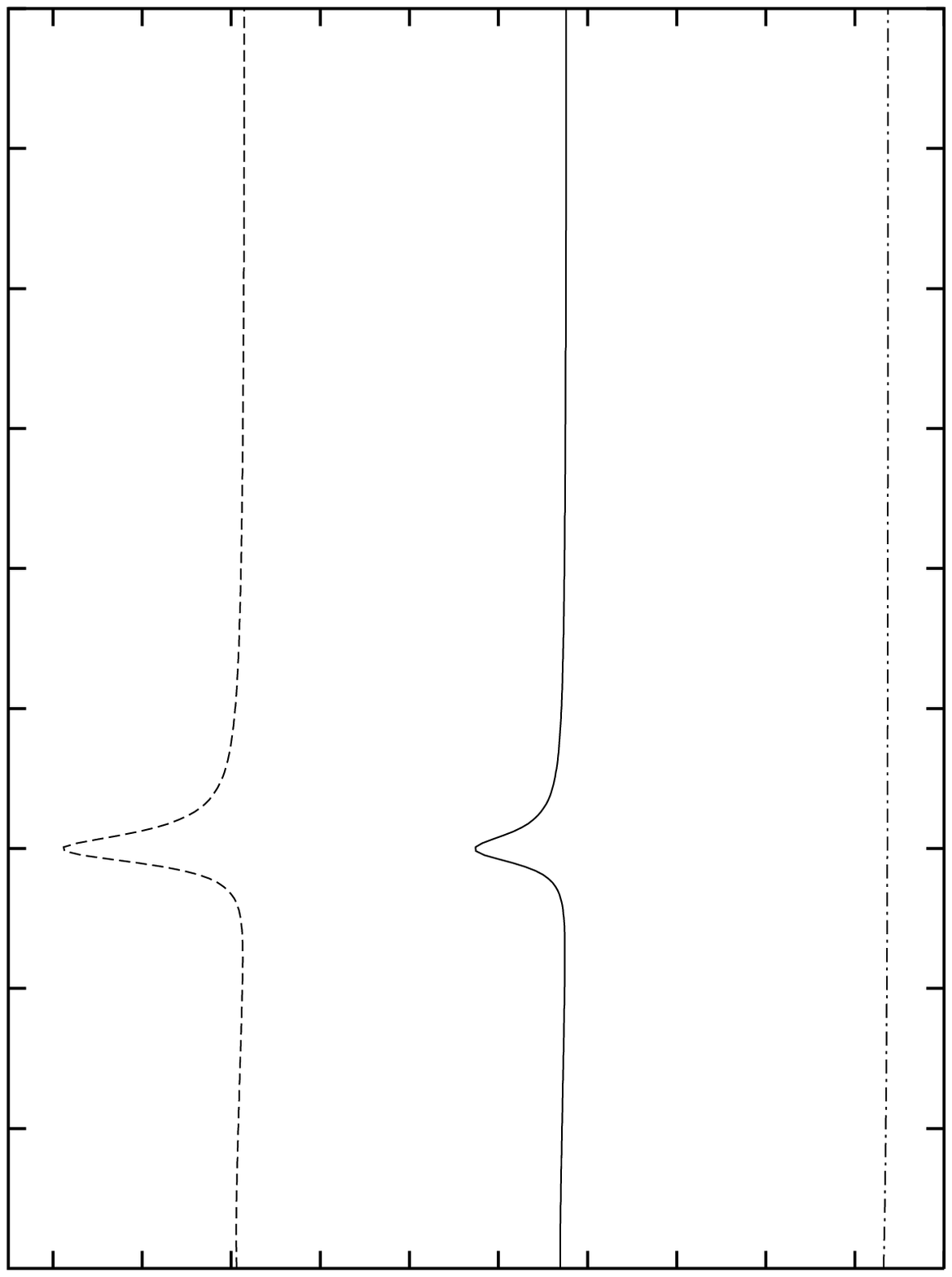}}}
\put(10,12){\begin{picture}(130,80)(0,0)
               \setlength{\unitlength}{1mm}
               \put(13,-5){$200$ }
               \put(45,-5){$400$ }
               \put(77,-5){$600$ }
               \put(109,-5){$800$ }
               \put(139,-5){$1000$ }
               \put(128.5,-10){$M_{\rm H}/$GeV}
               \put(-5,88){$\sigma$/pb }
               \put(-7,7.5){$0.4$}
               \put(-7,39){$0.8$}
               \put(-7,70.5){$1.2$}
               \end{picture}}
\end{picture}
\caption{Higgs-mass dependence of the \css\ in one-loop order for
$E_{\rm CMS} = 400\GeV$ ($0^\circ<\theta<180^\circ$)}
\label{fi:Higgsdep}
\end{figure}

Finally, we illustrate in \reffi{fi:Higgsdep} the
$\MH$-dependence of the
integrated \cs.
Outside the Higgs-resonance region the
variation is very weak, \ie
below $\sim1\%$.

\section{Summary}
\label{se:Sum}

The process $\AAtt$ will allow studies of the properties
of the top quark at high-energy photon--photon colliders
complementary to the ones achievable by hadron-hadron and
electron-positron colliders.
In particular, it is useful to investigate the electromagnetic couplings
of the top quark.

We have calculated the complete one-loop virtual and soft-photonic
radiative corrections to $\AAtt$ for arbitrarily polarized photons
in the electroweak Standard Model.
All so-called leading universal corrections -- such as the running
of the electromagnetic coupling, effects of the $\rho$-parameter or
leading QED logarithms -- are absent.
However, for equally polarized photons there are
large corrections close to the threshold of the process owing to
triangle top loops.

We have presented a detailed numerical analysis of the lowest-order
\css\ and the weak radiative corrections.
Below $1\TeV$ the weak corrections are in the range between 0\% and
$-10\%$.
The fermionic corrections vanish for opposite photon polarizations and
are otherwise small apart from the
triangle top-loop
contributions close to threshold. These dominate the corrections at
small energies.
At high energies, the bosonic corrections dominate
in particular for oppositely polarized photons.

Owing to the relatively high threshold the Higgs resonance is not very
pronounced. Outside the Higgs-resonance the dependence of the \cs\
on the Higgs-boson mass is very weak.
In particular, the corrections are finite in the limit of a large
Higgs-boson mass.

In summary, our results show that the weak $\Oa$ corrections to
$\AAtt$ are needed in order to match a per-cent level accuracy. For a
complete theoretical prediction QCD, QED and weak corrections as well
as finite-width effects of the top quark have to be combined.
Moreover, a convolution with a realistic spectrum for the incoming
photons has to be performed.

\appendix

\section*{Appendix}
\section{Interference of SME with the lowest-order matrix element}
\label{se:ConSME}

We list the results for the
non-vanishing
interference terms \refeq{eq:contractions}
for $t$-channel SME.
The corresponding results for the $u$-channel SME
are obtained by the following substitutions
\begin{eqnarray}
M^{\la_1\la_2,\{{\rm V,A}\},u}(i,t) &=& M^{\la_2\la_1,\{{\rm V,A}\},t}(i,u)
\Big\vert_{t \leftrightarrow u},
  \nn\\
M^{\la_1\la_2,\{{\rm V,A}\},u}(i,u) &=& M^{\la_2\la_1,\{{\rm V,A}\},t}(i,t)
\Big\vert_{t \leftrightarrow u}.
\label{eq:asmerel}
\end{eqnarray}

Equally polarized photons:
\begin{eqnarray}
M^{\pm\pm,{\rm V},t}(1,t) &=&
   8(t-\Mt^2)(u-\Mt^2)(3\Mt^4-3\Mt^2t+t^2-\Mt^2u)/s^2, \nn\\
M^{\pm\pm,{\rm V},t}(1,u) &=&
   8(t-\Mt^2)(u-\Mt^2)(\Mt^4-tu)/s^2, \nn\\
M^{\pm\pm,{\rm V},t}(2,t) &=&
   4(-2\Mt^6+2\Mt^4t-\Mt^2t^2+4\Mt^4u-4\Mt^2tu+2t^2u-\Mt^2u^2)/s, \nn\\
M^{\pm\pm,{\rm V},t}(2,u) &=&
   4(2\Mt^6-4\Mt^4t+\Mt^2t^2-2\Mt^4u+4\Mt^2tu+\Mt^2u^2-2tu^2)/s, \nn\\
M^{\pm\pm,{\rm V},t}(4,t) &=&
   4(\Mt^4-tu)(-4\Mt^4+3\Mt^2t-t^2+\Mt^2u+tu)/s^2, \nn\\
M^{\pm\pm,{\rm V},t}(4,u) &=&
   4(\Mt^4-tu)(-4\Mt^4+\Mt^2t+3\Mt^2u+tu-u^2)/s^2, \nn\\
M^{\pm\pm,{\rm V},t}(6,t) &=&
   2(\Mt^4-tu)(2\Mt^6-2\Mt^4t+\Mt^2t^2-4\Mt^4u+4\Mt^2tu-2t^2u+\Mt^2u^2)/s^2,
   \nn\\
M^{\pm\pm,{\rm V},t}(6,u) &=&
   2(\Mt^4-tu)(-2\Mt^6+4\Mt^4t-\Mt^2t^2+2\Mt^4u-4\Mt^2tu-\Mt^2u^2+2tu^2)/s^2,
   \nn\\
M^{\pm\pm,{\rm V},t}(11,t) &=&
   4\Mt^2(2\Mt^4+2\Mt^2s-s^2-4\Mt^2t+2t^2)/s,  \nn\\
M^{\pm\pm,{\rm V},t}(11,u) &=&
   8\Mt^2(\Mt^4-tu)/s, \nn\\
M^{\pm\pm,{\rm V},t}(12,t) &=&
   2\Mt^2(4\Mt^4-2\Mt^2t+t^2+2\Mt^2u-4tu-u^2)/s, \nn\\
M^{\pm\pm,{\rm V},t}(12,u) &=&
   2\Mt^2(4\Mt^4+2\Mt^2t-t^2-2\Mt^2u-4tu+u^2)/s, \nn\\
M^{\pm\pm,{\rm V},t}(16,t) &=&
   \Mt^2(\Mt^4-tu)(-4\Mt^4+2\Mt^2t-t^2-2\Mt^2u+4tu+u^2)/s^2, \nn\\
M^{\pm\pm,{\rm V},t}(16,u) &=&
   \Mt^2(\Mt^4-tu)(-4\Mt^4-2\Mt^2t+t^2+2\Mt^2u+4tu-u^2)/s^2.
\end{eqnarray}


All axial SME
for equally polarized photons
that do not violate CP-invariance vanish after contraction with
the Born matrix elements
\begin{equation}
M^{\pm \pm,{\rm A},t}(i,j) = 0, \quad i = 1, \ldots,17, \quad j = t,u.
\end{equation}

Oppositly polarized photons:
\begin{eqnarray}
M^{\pm\mp,{\rm V},t}(1,t) &=&
M^{\pm\mp,{\rm V},t}(1,u) = M^{\pm\mp,{\rm V},t}(4,t) =
M^{\pm\mp,{\rm V},t}(4,u) \nn\\ &=&
   4(\Mt^4-tu)(-6\Mt^4+4\Mt^2t-t^2+4\Mt^2u-u^2)/s^2, \nn\\
M^{\pm\mp,{\rm V},t}(6,t) &=&
M^{\pm\mp,{\rm V},t}(6,u) =
   2(t-u)(\Mt^4-tu)^2/s^2, \nn\\
M^{\pm\mp,{\rm V},t}(14,t) &=&
M^{\pm\mp,{\rm V},t}(14,u) =
   4\Mt^2(\Mt^4-tu), \nn\\
M^{\pm\mp,{\rm V},t}(16,t) &=&
M^{\pm\mp,{\rm V},t}(16,u) =
   -4\Mt^2(\Mt^4-tu)^2/s^2, \nn\\[1em]
M^{\pm\mp,{\rm A},t}(1,t) &=&
M^{\pm\mp,{\rm A},t}(1,u) = M^{\pm\mp,{\rm A},t}(4,t) =
M^{\pm\mp,{\rm A},t}(4,u) \nn\\ &=&
   \mp 4(t-u)(\Mt^4-tu)/s, \nn\\
M^{\pm\mp,{\rm A},t}(6,t) &=&
M^{\pm\mp,{\rm A},t}(6,u) =
   \pm 2(\Mt^4-tu)^2/s, \nn\\
M^{\pm\mp,{\rm A},t}(14,t) &=&
M^{\pm\mp,{\rm A},t}(14,u) =
   \pm 4\Mt^2(\Mt^4-tu).
\end{eqnarray}

\end{document}